\PassOptionsToPackage{dvipsnames}{xcolor}
\documentclass[aps,prb,reprint,superscriptaddress,showpacs]{revtex4-2}
\usepackage[T1]{fontenc}
\usepackage[utf8]{inputenc}
\usepackage[american]{babel}

\usepackage[ruled,vlined]{algorithm2e}
\usepackage{amsfonts}
\usepackage{amsmath}
\usepackage{amssymb}
\usepackage{bm}
\usepackage{booktabs}
\usepackage{gnuplot-lua-tikz}
\usepackage{graphicx}
\usepackage{hyperref}
\usepackage{latexsym}
\usepackage{mathtools}
\usepackage{subfig}
\usepackage{tikz}

\usetikzlibrary
  {arrows,calc,through,backgrounds,matrix,positioning,decorations.pathmorphing}

\def\XXint#1#2#3{{\setbox0=\hbox{$#1{#2#3}{\int}$ }
\vcenter{\hbox{$#2#3$ }}\kern-.6\wd0}}

\def\db{{\rm d}}
\def\vr{\boldsymbol r}

\newcommand{\e}{\varepsilon}
\newcommand{\te}{\underline{\e}}
\newcommand{\para}{\shortparallel}
\newcommand{\eff}{\varepsilon^\text{eff}}
\newcommand{\beff}{\beta^\text{eff}}
\newcommand{\keff}{k^\text{eff}}
\newcommand{\eb}{\tilde{\varepsilon}}
\newcommand{\teff}{\underline{\e}^\text{eff}}

\renewcommand{\vec}[1]{\boldsymbol{#1}}
\newcommand{\vE}{\vec E}

\newcommand{\vB}{\vec B}

\newcommand{\vk}{\vec k}

\newcommand{\ME}{\mathcal{E}}

\newcommand{\air}{{\text{air}}}
\newcommand{\sub}{{\text{sub}}}

\renewcommand{\Re}{\mathrm {Re}\,}
\renewcommand{\Im}{\mathrm {Im}\,}

\newcommand{\im}{\mathrm{i}}

\newcommand{\FTC}{\mathfrak{T}_{\text{c}}}

\begin{document}

\title{%
  Finite-size effects in wave transmission through plasmonic crystals:%
  \\ A tale of two scales}

\date{Draft of \today}

\author{Matthias Maier}
\thanks{\url{maier@math.tamu.edu}; \url{https://www.math.tamu.edu/~maier}}
\affiliation{Department of Mathematics, Texas A\&M University, College
  Station, Texas 77843, USA}

\author{Mitchell Luskin}
\affiliation{School of Mathematics, University of Minnesota, Minneapolis, %
  Minnesota 55455, USA}

\author{Dionisios Margetis}
\affiliation{Institute for Physical Science %
  and Technology, and Department of Mathematics, and Center for Scientific %
  Computation and Mathematical Modeling, University of Maryland, College %
  Park, Maryland 20742, USA.}

\begin{abstract}
  We study optical coefficients that characterize wave propagation through
  layered structures called plasmonic crystals. These consist of a
  \emph{finite number} of stacked metallic sheets embedded in dielectric
  hosts with a \emph{subwavelength spacing}. By adjustment of the
  frequency, spacing, number as well as geometry of the layers, these
  structures may exhibit appealing transmission properties in a range of
  frequencies from the terahertz to the mid-infrared regime. Our approach
  uses a blend of analytical and numerical methods for the distinct
  geometries with infinite, translation invariant, flat sheets and
  nanoribbons. We describe the transmission of plane waves through a
  plasmonic crystal in comparison to an \emph{effective dielectric slab} of
  equal total thickness that emerges from \emph{homogenization}, in the
  limit of zero interlayer spacing. We demonstrate numerically that the
  replacement of the discrete plasmonic crystal by its homogenized
  counterpart can accurately capture a transmission coefficient akin to the
  extinction spectrum, even for a relatively small number of layers. We
  point out the role of a geometry-dependent \emph{corrector field}, which
  expresses the effect of subwavelength surface plasmons. In particular, by
  use of the corrector we describe lateral resonances inherent to the
  nanoribbon geometry.
\end{abstract}

\maketitle

%%%%%%%%%%%%%%%%%%%%%%%%%%%%%%%%%%%%%%%%%%%%%%%%%%%%%%%%%%%%%%%%%%%%%%%%%%%%%%%%
%%%%%%%%%%%%%%%%%%%%%%%%%%%%%%%%%%%%%%%%%%%%%%%%%%%%%%%%%%%%%%%%%%%%%%%%%%%%%%%%
%%%%%%%%%%%%%%%%%%%%%%%%%%%%%%%%%%%%%%%%%%%%%%%%%%%%%%%%%%%%%%%%%%%%%%%%%%%%%%%%

\section{Introduction}

In the past few years, the advent of two-dimensional (2D) materials with
remarkable optoelectronic and thermal transport properties has
revolutionized several aspects of nanophotonics~\cite{Torres2014,
Geimetal2013, CastroNetoetal2009, Lowetal2017}. In particular, doped
monolayer graphene has an optical conductivity that allows this material to
interact strongly with light in a wide range of frequencies, from the
terahertz to the mid-infrared regime~\cite{Novoselovetal2012, LiBasov2008}.
This feature has inspired novel designs of plasmonic devices and
metamaterials with tunable optical properties~\cite{Atwateretal2018,
Daietal2015, Nemilentsau2016}.

Plasmonic crystals are a promising class of metamaterials. These structures
consist of stacked metallic layers which are arranged parallel to each
other with subwavelength spacing and are embedded in heterogeneous and
anisotropic dielectric hosts. By the tuning of the frequency, electronic
density, interlayer distance or number of layers, plasmonic crystals may
acquire unconventional optical properties~\cite{Mahmoodi2019, Dengetal2018,
Dengetal2015, Zhukovsky2014, Maier18}. To predict such properties, an
approach is to model the crystal as an \emph{effective continuous
medium}~\cite{Mahmoodi2019, Choy1999, Maier18}. This description may result
from a \emph{homogenization} procedure, in the asymptotic limit of
vanishing interlayer spacing~\cite{StuartPavliotis2007, Maier19b}. The
validity and implications of this simplified description for plasmonic
structures with a \emph{finite, possibly small,} number of layers are the
subjects of this paper. This focus distinguishes the present work from the
homogenization of periodic plasmonic structures of previous
treatments~\cite{Maier18,Maier19b}.

In recent experiments, the transmission properties of stacks consisting of
graphene sheets and insulators are measured or investigated at terahertz
frequencies; see,
e.g.,~\cite{Yanetal2012,Yaoetal2018,Ma2019,Nematpour2019}. A notable
outcome is that an increase in the number of layers, say, from one to five,
may cause a significant increase to the extinction spectrum of the
structure~\cite{Yanetal2012}. For such a small number of layers, it is
natural to wonder if the replacement of the inherently discrete, layered
structure by a suitably determined continuous dielectric medium can allow
for the accurate prediction of useful transmission properties. This issue
lies at the heart of modeling photonic heterostructures and metamaterials
of various geometries at the nanoscale~\cite{Geimetal2013,
KumarAvourisetal2015, HuAvourisetal2019, LeeAvourisetal2019, SiSun2017,
Huetal2018, Kimetal2019}.

In this paper, our goal is to address aspects of plasmonic metamaterials
that may be intimately connected to experiments. A novelty of our work is
that we provide an answer to the following question of practical appeal:
Can the homogenization procedure, which yields an effective continuous
medium, provide accurate predictions for wave transmission through
\emph{layered} plasmonic structures with \emph{prescribed} number of
conducting sheets and geometries of physical and technological interest?
Our approach is to apply homogenization theory in the (non-periodic)
setting of layered structures with \emph{finite} thickness. This view
enables us to study the effect of the number of layers, an experimentally
controllable parameter, on optical properties of practical importance.

First, we consider the prototypical setting with translation invariant,
planar graphene sheets intercalated between isotropic and homogeneous
dielectric hosts. For this configuration, we compute the Fresnel
coefficients analytically in closed forms via the transfer matrix approach
for transverse-magnetic (TM) polarization of the
fields~\cite{Yeh2005,Haus1984}. We also compare our findings to the
respective homogenization results for structures of finite total thickness,
adopting elements of a previous theory for periodic
structures~\cite{Mattheakisetal2016,Maier18}.

Second, we numerically examine the more realistic geometries with graphene
nanoribbons embedded in dielectric hosts. This study is carried out by the
following means: (i) the direct numerical computation of an appropriately
defined transmission coefficient via the finite element
method~\cite{Maier17}; and (ii) the application of homogenization, which
introduces the notion of the \emph{corrector field} to the leading order in
the interlayer spacing~\cite{Maier19b}. This field expresses the effect of
the surface plasmon-polariton (SPP), a subwavelength mode that can be
excited in 2D materials of suitably tuned conductivities. In the present
work, we combine the corrector field with the (finite) number of layers.
This approach renders our treatment and results distinct from those
in~\cite{Maier19b}. We show that the corrector field of the homogenized
structure can keep track of microscale \emph{lateral resonances} inherent
to the actual geometry of a single nanoribbon. This surface wave
interference effect is distinguished from \emph{interlayer resonances}. The
latter effect can also characterize the wave interaction between
translation invariant flat sheets at subwavelength spacing, when the
corrector tends to vanish. Our numerics indicate possible discrepancies
between predictions of the theory for the actual structure and the
corresponding effective model.

Our approach is motivated by the need to develop physical insight as well
as viable computational schemes for the design of complex nanophotonic
devices. In particular, the homogenization theory captures signatures of
microscale details in the form of weighted averages of material parameters
such as the permittivity of the dielectric host and the surface
conductivity of each sheet. In this framework, a structure that consists of
a finite number of conducting sheets at adjustable spacing, and their
dielectric hosts, is replaced by a suitably defined continuous medium (as
the spacing tends to zero) with the same total thickness. Hence, the number
of parameters and variables of the original problem is reduced in the
effective model. However, the weight for the requisite averages is in
principle provided by the corrector field which solves a boundary value
problem in the appropriately defined ``cell'' or ``representative volume
element''~\cite{Maier19b,StuartPavliotis2007}. This {\em cell problem} at
the microscale is a key ingredient of periodic homogenization. Our goal
here is to describe applications, advantages as well as possible
limitations of this simplified approach for the wave transmission through
realistic plasmonic structures.

Surprisingly, we find that the leading-order, low-energy homogenization
accurately captures the behavior of the relevant transmission coefficient
as a function of frequency even for a small, less than 10, number of
layers. In fact, the associated relative error can be negligible in
situations of possibly practical interest, and decreases as inverse
proportional to the number of layers (Secs.~\ref{sec:computation}
and~\ref{subsec:hom-accuracy}). The analytical and numerical computation of
discrete corrections to the homogenized result, due to the finite number of
layers (or, finite interlayer spacing) in the structure, for two selected
geometries and a wide range of frequencies is a highlight of our approach.
These ``finite-size''  effects exemplify the pivotal role of the corrector
field in the description of the effective dielectric medium if the sheets
(i.e., nanoribbons in our study) are not translation invariant in 2D. This
situation arises, for example, in the presence of edges or other defects on
each sheet. In this setting, the cell problem is characterized by plasmonic
lateral resonances, which we describe numerically. We reiterate that in the
special case of translation invariant layers the corrector field vanishes
identically. We compare our findings to previous theoretical models of
similar flavor for plasmonic crystals (Sec.~\ref{subsec:comparison}).

Notably, our theoretical results for translation invariant sheets here are
found in qualitative agreement with past experiments using stacks with
graphene sheets and insulators~\cite{Yanetal2012}. Furthermore, we point
out that our predictions on microscale resonances in nanoribbon
configurations can be experimentally testable (Sec.~\ref{subsec:expts}).
Bearing in mind possible practical considerations, we discuss the stability
of our homogenization results under random perturbations of parameters
(Sec~\ref{subsec:stability}).

Our work can be considered as an extension of previous studies that
focused on the ``epsilon-near-zero'' (ENZ) condition in plasmonic
heterostructures; see, e.g.,~\cite{SilveirinhaEngheta2006, Huangetal2011,
Moitraetal2013, LiMazuretal2015, Mattheakisetal2016, Maier18, Maier19a,
Maier19b}. One should recall that if the ENZ condition holds, it is
theoretically possible for a wave to propagate along a specified direction
of the crystal with almost no refraction at certain frequencies. In the
present paper, our results indicate that the ENZ condition is in principle
immaterial in the assessment of the accuracy of the homogenized
transmission coefficients. We note in passing that for a sufficiently large
number of layers, which is not the main focus of this paper, the ENZ
condition approximately characterizes the crossover between two distinct
behaviors of the wave transmission versus frequency and number of layers
(Sec.~\ref{subsec:hom-accuracy}). In seeking a connection of our
computations to previous studies in plasmonic
crystals~\cite{SilveirinhaEngheta2006, Huangetal2011, Moitraetal2013,
LiMazuretal2015, Mattheakisetal2016, Maier18, Maier19a, Maier19b}, we
analytically determine the behavior of the homogenized Fresnel coefficients
for a plasmonic slab in the parameter regime where the ENZ condition is
satisfied. In particular, we show how dissipation in the 2D material
affects the homogenized result for wave transmission in this regime.

It is tempting to compare our transfer matrix analysis for structures with
translation invariant sheets to the treatment of similar geometries for
hyperbolic metamaterials in~\cite{Mahmoodi2019}; see
also~\cite{Urbas2016,Galfsky2015,Cortesetal2012,Cortesetal2014-corrigendum}.
Here, we focus on the computation of Fresnel coefficients for wave
transmission through layered structures. Hence, we do not address the
dispersion relation of subwavelength (``high-$k$'') surface modes that can
be allowed by the layered structure~\cite{Mahmoodi2019}. In the Bloch wave
theory, such modes manifest as singularities (poles) in the complex plane
for the Fresnel coefficients as functions of the wave vector component
parallel to the sheets. The issue of how these singularities can be
computed accurately in a homogenized model lies beyond our present scope. A
related discussion, which touches upon plausible implications and
extensions of our results, can be found in Section~\ref{subsec:extensions}.

We should alert the reader about other questions that are left open in our
treatment. For example, we focus on the effects of TM-polarized fields, not
addressing the case with TE polarization. We do not study the effects that
a nonlocal conductivity of the 2D material, say, like the one caused by
viscous hydrodynamic electron flow~\cite{LucasFong2018}, may have on the
homogenization result. In our numerical computations, we adhere to the
treatment of structures with \emph{flat} conducting sheets; curved 2D
materials would have to be the subject of a separate study. As we allude to
above, our homogenization procedure is tailored to the treatment of
relatively low wavenumbers in the direction vertical to the sheets. The
modification of this procedure to take into account interlayer wave
phenomena at a length scale comparable to the subwavelength spacing is an
interesting direction of research.

The remainder of the paper is organized as follows. In
Sec.~\ref{sec:analysis}, we apply the transfer matrix approach to the
computation of the Fresnel coefficients for a structure with a finite
number of translation invariant, flat metallic sheets. In particular, we
derive the limit of these coefficients for small enough interlayer spacing.
Section~\ref{sec:enz} focuses on the description of the Fresnel
coefficients under the ENZ condition. In Sec.~\ref{sec:homogen-reson}, we
revisit the general homogenization framework for periodic layered
structures, particularly the emergent corrector field; and then adopt the
resulting effective permittivity for a plasmonic structure of finite
thickness. Section~\ref{sec:computation} provides numerical results for
plasmonic crystals in the distinct cases with translation invariant sheets
and nanoribbons in comparison to their homogenized counterparts. In
Sec.~\ref{sec:discussion}, we discuss implications and extensions of our
results. Section~\ref{sec:conclusion} concludes the paper with a summary of
the results. The appendices provide technical yet non-essential derivations
as well as a review of the general homogenization theory~\cite{Maier19b}.
The time dependence is $e^{-\im \omega t}$ throughout ($\omega$ is the
angular frequency).

%%%%%%%%%%%%%%%%%%%%%%%%%%%%%%%%%%%%%%%%%%%%%%%%%%%%%%%%%%%%%%%%%%%%%%%%%%%%%%%%
%%%%%%%%%%%%%%%%%%%%%%%%%%%%%%%%%%%%%%%%%%%%%%%%%%%%%%%%%%%%%%%%%%%%%%%%%%%%%%%%
%%%%%%%%%%%%%%%%%%%%%%%%%%%%%%%%%%%%%%%%%%%%%%%%%%%%%%%%%%%%%%%%%%%%%%%%%%%%%%%%

\section{Transfer matrix approach: Translation invariant sheets}
\label{sec:analysis}

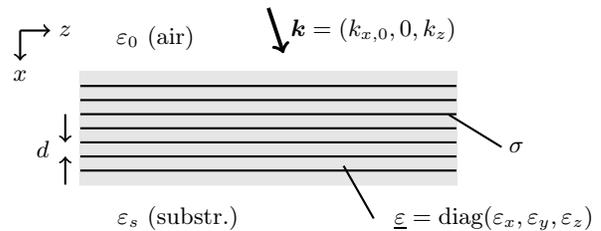
\begin{figure}
  \centering
    \begin{tikzpicture}
      \def\nlayers{8}
      \def\nrepeti{19}
      \def\delta{1.5 / 4. / \nlayers};
      \path[thick, draw=black!10, fill=black!10]
        (-2.5, -0.75) -- (2.5, -0.75) -- (2.5, 0.75) -- (-2.5, 0.75) -- cycle;
       \foreach \m in {1,...,7} {
           \def\y{-0.75 + 4. * \delta * \m};
           \draw[thick] (-2.5, \y) -- (2.5, \y);
      }
      \path [thick, ->, draw] (-3.3, 1.3) -- (-2.9, 1.3);
      \path [thick, ->, draw] (-3.3, 1.3) -- (-3.3, 0.9);
      \node at (-3.3, 0.7) {$x$};
      \node at (-2.7, 1.3) {$z$};

      \path [thick, ->, draw] (-2.7, -0.75) -- (-2.7, -0.75+4. * \delta * 2);
      \path [thick, ->, draw] (-2.7, -0.75 + 4.*\delta*5) -- (-2.7, -0.75+4. * \delta * 3);
      \node at (-3.0, -0.75 + 4.*\delta*2.5) {$d$};

      \path [line width=0.5mm, ->, draw] (0, 1.60) -- (0.2, 1.00);
      \node at ( 1.4,  1.3) {$\vec k=(k_{x,0},0,k_z)$};
      \node at (-1.5,  1.2) {$\varepsilon_0$ (air)};
      \node at (-1.2, -1.2) {$\varepsilon_s$ (substr.)};
      \node at (3.0, -1.2)
      {$\underline\varepsilon=\text{diag}(\e_x,\e_y,\e_z)$};
      \path [thick, draw] (1.4, -1.2) -- (1.0, -0.5);
      \node at (3.3, -0.25) {$\sigma$};
      \path [thick, draw] (3.1, -0.25) -- (2.40, 0.18);
    \end{tikzpicture}
  \caption{Problem geometry. The layered structure has interlayer spacing
    $d$. The conducting sheets are isotropic with conductivity $\sigma$;
    and the dielectric host has tensor permittivity
    $\te=\text{diag}(\e_\perp,\e_\para,\e_\para)$ in the indicated
    coordinate system. The structure lies between air (permittivity $\e_0$)
    and substrate (permittivity $\e_{\text{s}})$.  All materials have
    uniform permeability $\mu$. A plane wave (wave vector $\boldsymbol k$)
    is incident upon the structure from air.}
  \label{fig:sandwich}
\end{figure}

In this section, we apply the transfer matrix
formalism~\cite{Yeh2005,Haus1984} to the calculation of the reflection and
transmission coefficients for a plasmonic crystal with finite thickness
under TM polarization of the electromagnetic field. The crystal consists of
a finite number of infinitely extended sheets embedded in dielectric hosts,
as shown in Fig.~\ref{fig:sandwich}. The interlayer spacing of the
structure is $d$, and the number of conducting sheets is $(N-1)$; thus, the
total thickness is $H=Nd$. Each dielectric host and conducting sheet is
translation invariant in $y$ and $z$ (but not in $x$). This geometry should
be contrasted to the case with nanoribbons which is studied numerically in
Sec.~\ref{sec:computation} (see also
Sec.~\ref{subsec:hom-Fresnel-diagonal_e} for the respective Fresnel
coefficients).

We assume that each sheet has the homogeneous and isotropic surface
conductivity $\sigma(\omega)$. The dielectric host can be anisotropic with
permittivity represented by the spatially constant matrix
\begin{align*}
  \te(\omega)=\text{diag}(\e_x(\omega),\e_{y}(\omega),\e_{z}(\omega))
\end{align*}
where in general $\e_\ell$ are distinct ($\ell=x,\,y,\,z$). This
anisotropic permittivity is necessary for the study at hand, particularly
the computations of Sec~\ref{sec:computation}. Our motivation for choosing
this model for $\te$ is twofold.  First, anisotropic dielectric materials
such as the hexagonal boron nitride (hBN) are of theoretical and
experimental interest in plasmonics~\cite{Daietal2015,Nemilentsau2016}.
Second, the Fresnel coefficients derived in this section, in an anisotropic
setting, are used in Sec.~\ref{sec:homogen-reson} when the discrete system
is replaced by an effective continuous medium (which is a single,
macroscopically thick slab). In that case, the anisotropy in the effective
description emerges from homogenization.

The layered structure lies in the region with $0<x<H$ between two
unbounded, uniform dielectrics that have scalar permittivities $\e_0$ (for
air) and $\e_{\text{s}}$ (substrate). Thus, the free space and substrate
occupy the regions with $x<0$ and $x>H$, respectively. The magnetic
permeability is equal to $\mu$ in all media.

%%%%%%%%%%%%%%%%%%%%%%%%%%%%%%%%%%%%%%%%%%%%%%%%%%%%%%%%%%%%%%%%%%%%%%%%%%%%%%%%

\subsection{Formulation}
\label{subsec:formulation}

Assuming that the electromagnetic field $(\vE,\vB)$ is TM polarized, we set
$\vE=(E_x,0,E_z)$ and $\vB=(0,B_y,0)$ where all field components are
$y$-independent. Suppose that a plane wave is incident upon the layered
structure from the air, for $x<0$. This wave is partially reflected from
and transmitted through the structure. In our configuration
(Fig.~\ref{fig:sandwich}), the $z$-component, $k_z$, of the wave vector,
$\vk$, of any associated plane wave is a prescribed invariant of the
problem. In view of the fixed polarization, we can thus reduce the boundary
value problem for Maxwell's equations for $(\vE,\vB)$ to the transmission
problem for a single field component, e.g., the $z$-component, $E_z$, of
$\vE$~\cite{Maier18,Maier17}.

Hence, we aim to compute the related reflection and transmission
coefficients by the following procedure. For every dielectric slab of the
configuration, we write
\begin{align*}
  E_z(x,z) = \ME(x)\,e^{ik_z\,z},
  \quad
  \ME(x) = A\,e^{-\im k_x\,x}\;+\;C\,e^{\im k_x\,x}
\end{align*}
where ${\rm Re}\, k_x>0$ for definiteness. In the dielectric host, the
wavenumber $k_x$ is found to be
\begin{equation}
  \label{eq:kx-beta-disp}
  k_x
  =
  \sqrt{\frac{\e_\para}{\e_\perp}\big(k_\perp^2-k_z^2\big)}=\beta(k_z),\quad
  k_\perp^2=\omega^2\mu\e_\perp
\end{equation}
where $\e_\perp=\e_x$ and $\e_\para=\e_z$; see
Appendix~\ref{app:transfer_matrix} for a derivation of dispersion
relation~\eqref{eq:kx-beta-disp} for $k_x$. For ease in notation, we
henceforth denote the $k_x$ in the host slab by $\beta$. Note that in the
present case with TM polarization, the ``lateral'' matrix element $\e_y$ of
the diagonal $\te$ becomes irrelevant; thus, we could have used the
permittivity matrix $\te=\text{diag}(\e_\perp,\e_\para,\e_\para)$ without
loss of generality.

The amplitudes $A$ and $C$ entering $\ME(x)$ depend on the corresponding
medium and layer, and can in principle be determined from the requisite
transmission conditions, as outlined below. In particular, in air ($x<0$)
the field component $E_z$ is expressed as
\begin{align*}
  E^\air_z = \ME^\air(x)\,e^{ik_z\,z},
  \quad
  \ME^\air(x) = e^{\im k_{x,0}\,x}\;+\;R\,e^{-\im k_{x,0}\,x}
\end{align*}
where $k_{x,0}=\sqrt{k_0^2-k_z^2}$ and $R(k_z)$ is the reflection
coefficient with  $k_0^2=\omega^2\mu\varepsilon_0$. In the substrate
($x>H$), the solution for $E_z(x,z)$ becomes
\begin{align*}
  E^\sub = \ME^\sub(x)\,e^{ik_z\,z},
  \quad
  \ME^\sub(x) = T\,e^{\im k_{x,\text{s}}\,x}
\end{align*}
where $T(k_z)$ is the transmission coefficient and
$k_{x,\text{s}}=\sqrt{k_\text{s}^2-k_z^2}$ with
$k_{\text{s}}^2=\omega^2\mu\varepsilon_{\text{s}}$.

The task at hand is to determine $R(k_z)$ and $T(k_z)$ explicitly. We thus
apply the necessary boundary conditions through the dielectric interfaces
and conducting sheets; see Appendix~\ref{app:transfer_matrix} for details.
First, we impose continuity of $E_z(x,z)$ across each conducting sheet as
well as across the interfaces of the dielectric host with air or substrate;
thus, $\ME(x)$ must be continuous at $x=nd$ for $n=0,\,1,\,\ldots
N$~\cite{Maier17}. Second, we require that the nonzero tangential ($y$-)
component of $\vB$, which is proportional to $d\ME/dx$, be continuous
across the dielectric interfaces (at $x=0, H$). In addition, this component
must experience a \emph{jump} equal to the surface current, $\sigma E_z$,
across each conducting sheet (at $x=nd$ for $n=1,\,\ldots,
N-1$)~\cite{Maier17}.

%%%%%%%%%%%%%%%%%%%%%%%%%%%%%%%%%%%%%%%%%%%%%%%%%%%%%%%%%%%%%%%%%%%%%%%%%%%%%%%%

\subsection{Formulas for Fresnel coefficients}
\label{subsec:eval-Fresnel}

Consider the layered structure of Fig.~\ref{fig:sandwich}. We now address
the full transmission problem by the transfer matrix
approach~\cite{Yeh2005,Haus1984}. Here, we briefly outline the procedure,
and state the main results for the Fresnel coefficients $R(k_z)$ and
$T(k_z)$. Details of the derivation can be found in
Appendix~\ref{app:transfer_matrix}.

The main idea is to view the whole transmission problem as a cascade of
elementary propagation problems. We then connect the amplitudes of
$E_z(x,z)$ for $x<0$ and $x>H$ via the multiplication of the constituent
transfer matrices. In this vein, we define the following matrices:
\begin{align*}
  \mathcal{T}_{\text{I}} &=
  \begin{pmatrix}
    e^{-i\beta d} & 0 \\
    0 & e^{i\beta d}
  \end{pmatrix}=\mathcal{T}_{\text{I}}(d),\\
  \mathcal{T}_{\text{II}} &=
  \begin{pmatrix}
    1-\frac{id}{2\beta}\big[(\beta^{\text{eff}})^2 - \beta^2\big]
    & -\frac{id}{2\beta}\big[(\beta^{\text{eff}})^2 - \beta^2\big]
    \\
    \frac{id}{2\beta}\big[(\beta^{\text{eff}})^2 - \beta^2\big]
    & 1+\frac{id}{2\beta}\big[(\beta^{\text{eff}})^2 - \beta^2\big]
  \end{pmatrix}
\end{align*}
where $\beff$ denotes the \emph{effective} wavenumber
\begin{align}\label{eq:beta-eff-def}
  \beff(k_z) =\sqrt{\beta(k_z)^2 +\im\,
  \frac{\omega\mu\sigma}{d}\, \frac{k^2_\perp - k_z^2} {k^2_\perp}}.
\end{align}
In the above, $\mathcal T_{\text{I}}$ characterizes the propagation of the
$z$-directed electric field, $\ME(x)$, in the dielectric host by distance
$d$.  The matrix $\mathcal T_{\text{II}}$ describes the transmission of
$\ME(x)$ through a sheet of surface conductivity $\sigma$ that is immersed
in the dielectric host at position $x=0$ (see
Appendix~\ref{app:transfer_matrix}). The effective wavenumber $\beff$ is
introduced in hindsight, for later algebraic convenience: This definition
serves our purpose of homogenizing the structure in the limit $\beta d\to
0$ by keeping certain nondimensional parameters fixed. The role of this
$\beff$ becomes more clear in Sec.~\ref{sec:small_spacing}. In particular,
we will require that $\beff H$ is independent of $d$ as $\beta d\to 0$.
This requirement is equivalent to the statement that $\sigma$ scales
linearly with $d$ (or $\sigma\diagup d\simeq {\rm const.}$) if the other
material parameters are considered as fixed~\cite{Maier19b}. The physical
significance of this scaling of $\sigma$ with $d$ is discussed in
Sec.~\ref{sec:small_spacing}.

By the transfer matrix approach, the amplitude vector $(A,C)$ for $\ME(x)$
in the first slab with $0<x<d$ is connected to the respective amplitudes in
the last slab where $(N-1)d<x<H$. This is carried out by the successive
application of $\mathcal T_{\text{I}}(nd)$ and $\mathcal{T}_{\text{II}}$
through the layered structure (for $n=1,\,\ldots\,,N-1$). Subsequently, we
have to relate the field amplitudes in the above extremal slabs to the
pairs $(R,1)$ and $(T,0)$, in the air and substrate  (see
Fig.~\ref{fig:sandwich}). To this end, we impose the continuity of $E_z$
and $B_y$ at $x=0$ and $x=H$.

After some algebra, we arrive at the following closed-form expressions for
the reflection and transmission coefficients, $R$ and $T$ (see
Appendix~\ref{app:transfer_matrix}):
\begin{align}
  \begin{cases}
    \begin{aligned}
      R\;&=\;-\;\frac%
      {\eb_{\text{s}+}\,t_{1-} -\eb_{\text{s}-}\,t_{2-}\,e^{\im 2\beta d}}
      {\eb_{\text{s}+}\,t_{1+}-\eb_{\text{s}-}\,t_{2+}\,e^{\im2\beta d}},
      \\[0.75em]
      T\;&=\;e^{-\im k_{x,\text{s}}H}e^{\im\beta d}\,
      \frac{t_{1+}\,t_{2-}-t_{2+}t_{1-}}
      {\eb_{\text{s}+}\,t_{1+}-\eb_{\text{s}-}\,t_{2+}\,e^{\im2\beta d}}.
    \end{aligned}
  \end{cases}
  \label{eq:fresnel}
\end{align}
Here, we define the matrix elements $t_{ij}$ ($i,\,j=1,\,2$) by
\begin{align*}
  \begin{pmatrix}
    t_{11} & t_{12} \\ t_{21} & t_{22}
  \end{pmatrix}
  \;=\; \Big(\mathcal{T}_{\text{II}}\mathcal{T}_{\text{I}}\Big)^{N-1}~;
\end{align*}
and also introduce the parameters
\begin{equation*}
  \eb_{\ell\pm}\;=\;1\,\pm\,\frac{\e_{\ell}}{\e_\perp}
  \frac{k_\perp^2-k_z^2}{\beta\,k_{x,\ell}},
  \quad\text{for }\ell=0,\,\text{s}~,
\end{equation*}
\begin{equation*}
  t_{i\pm}=t_{i1}\,\eb_{0\pm}+t_{i2}\,\eb_{0\mp}\quad (i=1,\,2).
\end{equation*}
We note in passing that the analytical computation of the matrix elements
$t_{ij}$ can be carried out via the diagonalization of
$\mathcal{T}_{\text{II}}\mathcal{T}_{\text{I}}$; see
Appendices~\ref{app:transfer_matrix}
and~\ref{app:transfer_matrix_expansion}. Specifically, by writing
$\mathcal{T}_{\text{II}}\mathcal{T}_{\text{I}}=\mathcal
S\,\text{diag}(\lambda_+,\lambda_-)\mathcal S^{-1}$ where $\mathcal S$ is a
non-singular matrix and $\lambda_\pm$ are the eigenvalues of
$\mathcal{T}_{\text{II}}\mathcal{T}_{\text{I}}$, we apply the identity
\begin{align*}
  \begin{pmatrix}
    t_{11} & t_{12} \\ t_{21} & t_{22}
  \end{pmatrix}
  \;=\; \mathcal
  S\,\text{diag}\big(\lambda_+^{N-1},\lambda_-^{N-1}\bigr)\,\mathcal
  S^{-1}.
\end{align*}
Explicit formulas for $\mathcal S$ and $\lambda_\pm$ are given in
Appendix~\ref{app:transfer_matrix_expansion}.

%%%%%%%%%%%%%%%%%%%%%%%%%%%%%%%%%%%%%%%%%%%%%%%%%%%%%%%%%%%%%%%%%%%%%%%%%%%%%%%%

\subsection{Limit of small interlayer spacing ($\beta d\to 0$)}
\label{sec:small_spacing}

Next, we illustrate the derivation of approximate formulas for the
coefficients $R$ and $T$ in the limit as the number of layers, $N$, is
sufficiently large ($N\gg 1$) and the interlayer spacing, $d$, is small
enough. We discuss the underlying key assumptions in detail. In our
limiting process the parameter $\beta H=N\beta d$ and the ratio
$\sigma\diagup d$ are kept fixed; cf.~\cite{Maier19b}. The resulting limit
formally expresses the homogenization of plasmonic crystals of finite
thickness.

Specifically, a set of assumptions is described by~\cite{Maier19b}
\begin{align*}
  \big|\beta d\big|\ll 1,
  \qquad
  \left|\frac{\omega\mu\sigma(\omega)}{\beta}\right|\ll 1,
\end{align*}
with fixed $\beta H$. The first condition ($|\beta d|\ll 1$) means that $d$
is small compared to the wavelength of propagation in the $x$-direction
inside the dielectric host. The second condition ($|\omega\mu\sigma\diagup
\beta|\ll 1$) implies the subwavelength character of the TM-polarized SPP
associated with an {\em isolated} conducting sheet in the (unbounded)
dielectric host. This condition expresses the separation of two distinct
length scales: One scale is related to the wavenumber $\beta$ describing
the $x$-directed propagation in the dielectric slab; and another is related
to the SPP wavenumber, which scales as $1/\sigma$ on the conducting
sheet~\cite{Maier18,Maier19b}. Note that TM-polarized SPPs cannot be
excited on a single sheet by plane waves in the present geometry.
Nonetheless, the above interpretation in terms of a scale separation helps
us to point out the role of the spacing $d$, as we explain next.

In regard to the spacing $d$, we {\em additionally} assume that
\begin{align*}
\frac{\omega\mu\sigma(\omega)}{\beta^2 d} \simeq {\rm const}.
\end{align*}
which means that the length $|\omega\mu\sigma(\omega)\diagup \beta^2|$
scales linearly with $d$ as $\beta d\to 0$. By Eq.~\eqref{eq:beta-eff-def},
this assumption is compatible with $\beff H$ being kept fixed, independent
of $d$. This choice of scaling for $\omega\mu\sigma(\omega)\diagup \beta^2$
here implies that the wavelength of the TM-polarized SPP associated with
the isolated sheet becomes comparable to the interlayer spacing. Hence, in
the limit $\beta d\to 0$ the strength of the coupling of possible SPPs, or
surface plasmonic modes, on neighboring sheets is nearly constant. The
dimensionless parameter $\omega\mu\sigma\diagup (\beta^2 d)$ expresses the
strength of this coupling.

An alternative way to state the above scaling is to write
\begin{equation*}
  \sigma\diagup d \simeq \text{const}.,
\end{equation*}
if the material parameters other than $\sigma$ are considered as fixed,
independent of $d$~\cite{Maier19b}. This choice of scaling $\sigma$
linearly with $d$ implies that the total surface current on the sheets
remains \emph{finite} in the limiting process ($\beta d\to 0$).

Without sacrificing the essential physics of the problem, for the sake of
simplicity we set $\e_{\text{s}}=\e_0$. In other words, we assume that the
whole layered structure is immersed in a homogeneous and isotropic medium
(air). Consider normal incidence of the incoming plane wave, i.e., take
$k_z=0$.

Let us now turn our attention to the exact formulas of
Eq.~\eqref{eq:fresnel}. By expanding the Fresnel coefficients in powers of
$\beta d$, we obtain the following results to the leading order in $\beta
d$ (see Appendix~\ref{app:transfer_matrix_expansion}):
\begin{align}
  \label{eq:homogenized_layers}
  \begin{cases}
    \begin{aligned}
      R\simeq R^{\text{eff}}&=-
      \frac%
      {\big[(\beff)^2-k_0^2\big]\,\tan(\beff H)}
      {\big[(\beff)^2+k_0^2\big]\,\tan(\beff H)+2\,\im k_0\beff},
      \\[0.5em]
      T\simeq T^{\text{eff}}&=
      \frac
      {e^{-\im k_0H}\,2\,\im k_0\beff\sec(\beff H)}
      {\big[(\beff)^2+k_0^2\big]\,\tan(\beff H)+2\,\im k_0\beff}.
    \end{aligned}
  \end{cases}
\end{align}
These equations define the homogenized Fresnel coefficients for the layered
medium of total thickness $H$.
Equation~\eqref{eq:homogenized_layers} can be extended to the case with
$k_z\neq0$, i.e., oblique incidence of the plane wave from air. This
extension in the effective Fresnel coefficients can be carried out by
replacing $k_0$ by the wavenumber
\begin{align*}
  \tilde k_0 \;=\; k_0\;\frac{k_0}{k_{x,0}}\;\frac{k^2_\perp-k_z^2}{k^2_\perp}.
\end{align*}

A few remarks on the above approximate formulas for $R$ and $T$ are in
order. These coefficients suggest that the layered structure is effectively
replaced by a continuous medium characterized by the wavenumber $\beff$ for
propagation in the transverse ($x$-) direction. By
Eq.~\eqref{eq:beta-eff-def} with $k_z=0$, we have
\begin{align*}
  \beff=\beff(0) = \sqrt{k_\para^2 +\im\,
  \frac{\omega\mu\sigma}{d}};\quad k_\para=\omega\sqrt{\mu\e_\para}.
\end{align*}
Accordingly, in this limit, the wave in the plasmonic structure encounters
the effective dielectric permittivity $\te^{\text{eff}}={\rm
diag}(\e_\perp^{\rm eff}, \e_\para^{\rm eff}, \e_\para^{\rm eff})$,
where $\e_\perp^{\rm eff}=\e_\perp$ and
\begin{align*}
  \e_\para^{\text{eff}}=\e_\para+\im\,\frac{\sigma(\omega)}{\omega d}~.
\end{align*}
The emergent anisotropy of the effective dielectric permittivity,
$\te^{\text{eff}}$, is intrinsic to the structure geometry: As $\beta d\to
0$, the surface conductivities of individual sheets conspire to give rise
to a bulk property (volume conductivity) that necessarily modifies only the
\emph{lateral} matrix elements of $\te$ in the effective-medium
decsription. More generally, geometric asymmetries between the transverse
and lateral directions, relative to each layer, at the scale of the
interlayer spacing, $d$, are expected to give rise to material anisotropy
in the homogenization limit~\cite{Maier19b,StuartPavliotis2007}.

The results of this section are compatible  with the dispersion relation
derived via Bloch wave theory in~\cite{Mattheakisetal2016,Maier18} for a
periodic array of conducting sheets. Recall that Eq.~\eqref{eq:fresnel} is
valid for infinite, translation invariant layers. In this case, there are
no (lateral) plasmonic resonances inherent to the geometry of the isolated
2D material. In contrast, each nanoribbon is characterized by resonances
related to the strip width. These subtle effects of geometry are captured
by the corrector field~\cite{Maier19b}. Regarding the general
homogenization theory and the corrector field, the interested reader is
referred to Sec.~\ref{sec:homogen-reson}.

Numerical comparisons of a transmission property (``complementary
transmission spectrum'') related to the homogenized coefficients
\eqref{eq:homogenized_layers} to the exactly computed formula based on
Eq.~\eqref{eq:fresnel} indicate that the homogenization results are
reasonably accurate. The accuracy persists even for a relatively small
number, $N$, of layers. A quantitative study of this issue for the
practically appealing cases~\cite{Yanetal2012} with $N=4$, $8$ and $16$ is
presented in Sec.~\ref{sec:computation}.

%%%%%%%%%%%%%%%%%%%%%%%%%%%%%%%%%%%%%%%%%%%%%%%%%%%%%%%%%%%%%%%%%%%%%%%%%%%%%%%%
%%%%%%%%%%%%%%%%%%%%%%%%%%%%%%%%%%%%%%%%%%%%%%%%%%%%%%%%%%%%%%%%%%%%%%%%%%%%%%%%
%%%%%%%%%%%%%%%%%%%%%%%%%%%%%%%%%%%%%%%%%%%%%%%%%%%%%%%%%%%%%%%%%%%%%%%%%%%%%%%%

\section{Transmission at the ENZ condition}
\label{sec:enz}

In this section, we describe the effect of a generalized ENZ condition on
the homogenized coefficients of Eq.~\eqref{eq:homogenized_layers}. In our
setting (Fig.~\ref{fig:sandwich}), the idea underlying this condition is
suggested by the observation that a wave propagating in the effective
medium of the plasmonic crystal in the ($x$-) direction transverse to the
layers can be suitably tuned to experience almost no phase
delay~\cite{Mattheakisetal2016,Maier18}.

This theoretical possibility is typically introduced for periodic plasmonic
crystals without ohmic losses; see, e.g,~\cite{Mattheakisetal2016}. Our
study here offers an extension of this concept to include
finite-number-of-layers and dissipation effects. The possible role of the
ENZ condition in the description of wave transmission through a plasmonic
structure with a finite number of layers for a wide range of frequencies is
discussed in Sec.~\ref{subsec:hom-accuracy}.

First, we review the concept of the ENZ condition in the absence of
dissipation (for $\Re\sigma\simeq 0$). A means of arriving at the ENZ
condition is to require that the effective wavenumber of
Eq.~\eqref{eq:beta-eff-def} vanishes for any given $k_z$. Alternatively, at
least one (real) eigenvalue of the effective permittivity tensor becomes
zero~\cite{Maier19b}; here, this requirement yields
$\e_\para^{\text{eff}}=0$. For periodic plasmonic crystals, this condition
entails that a branch of the dispersion relation $k_x(k_z)$ for the layered
structure approaches a Dirac cone near the center of the Brillouin
zone~\cite{Mattheakisetal2016,Maier18}.

More generally, if the dielectric media are lossless but each conducting
sheet is {\em dissipative} ($\Re\sigma>0$), we define the ENZ condition by
\begin{align}\label{eq:ENZ-dissip}
  \Re\e_\para^{\text{eff}}=0
  \;\Rightarrow\;
  \Re\Big\{\e_\para\;+\im\;\frac{\sigma(\omega)}{\omega d}\Big\}\;=\;0.
\end{align}
This equation entails $d=(\Im\sigma)/(\omega\e_\para)$.
Assuming again normal incidence of the incoming plane wave, i.\,e.,
$k_z=0$, as well as $\e_{\text{s}}=\e_0$,
by Eqs.~\eqref{eq:beta-eff-def} and~\eqref{eq:ENZ-dissip} we
obtain
\begin{align*}
  [\beff(0)]^2 = \frac{\im\omega\mu(\Re\sigma)}{d}\,\Rightarrow\,
  \beff = \beff(0)=\gamma\,k_\para
\end{align*}
where $k_\para=\omega\sqrt{\mu\e_\para}= \omega\sqrt{\mu\e}$ and
\begin{align*}
  \gamma=e^{\im\pi/4}\,k_\para^{-1}\sqrt{\frac{\omega\mu(\Re\sigma)}{d}}
  =(1+\im)\sqrt{\frac{\Re\sigma}{2\,\Im\sigma}}.
\end{align*}
In the above equation, we replaced $d$ by $(\Im\sigma)/(\omega\e_\para)$
according to the ENZ condition. The dimensionless complex parameter
$\gamma$ measures the effect of dissipation; and indicates the deviation of
condition~\eqref{eq:ENZ-dissip} from its dissipation-free
counterpart~\cite{Mattheakisetal2016,Maier18}. Typically, for a range of
terahertz frequencies in doped monolayer graphene, we expect that
$|\gamma|\ll 1$~\cite{Lowetal2017}.  Note that the symbol $\gamma$ here
should not be confused with the same symbol used typically to denote the
(positive) damping figure of merit for SPPs~\cite{Lowetal2017}.
Qualitatively, however, both quantities express the effect of dissipation;
more precisely, $-\im\gamma^2$ equals the standard SPP damping
ratio~\cite{Lowetal2017}.

We proceed to provide simplified formulas for $R^{\text{eff}}$ and
$T^{\text{eff}}$ when condition~\eqref{eq:ENZ-dissip} holds.
Equation~\eqref{eq:homogenized_layers} yields
\begin{align*}
  \begin{cases}
    \begin{aligned}
      R^{\text{eff}}\to R_{\text{ENZ}} &=
      \frac%
      {(k_0^2-\gamma^2 k_\para^2)\tan(\gamma k_\para H)}
      {(k_0^2+\gamma^2k_\para^2)\tan(\gamma k_\para H)+2\im \gamma k_\para k_0}\,,
      \\[0.5em]
      T^{\text{eff}}\to T_{\text{ENZ}}&=
      \frac
      {e^{-\im k_0H}\,2\,\im \gamma k_0k_\para \sec(\gamma k_\para H)}
      {\big(k_0^2+\gamma^2 k_\para^2\big)\tan(\gamma k_\para H)+2\,\im
      \gamma k_0k_\para}.
    \end{aligned}
  \end{cases}
\end{align*}

It is reasonable to assume that $k_\para H$ is fixed and not large (see
Sec.~\ref{sec:small_spacing}). Thus, the smallness of $|\gamma|$ in
\emph{weakly dissipative} 2D materials should imply that $| \gamma k_\para
H|\ll 1$. Accordingly, we can expand the coefficients $R_{\text{ENZ}}$ and
$T_{\text{ENZ}}$ in powers of the parameter $\gamma k_\para H$. By
expanding up to second order in this parameter, after some algebra we
obtain
\begin{multline*}
  R_{\text{ENZ}} \;\simeq\; \frac{k_0 H}{2\,\im+k_0H}
  \\
  \;+\;
  \frac{\im\gamma^2}{(2\,\im+k_0H)^2}
  \,\cdot\Bigg\{
  \frac{2}{3}\;-\;
  \frac{2+2\im\,k_0H}{(k_0 H)^2}
  \bigg\}
  \,(k_0 H)(k_\para H)^2,
\end{multline*}
and
\begin{multline*}
  e^{\im k_0H}\,T_{\text{ENZ}} \;\approx\;
  \frac{2\,\im}{2\,\im+k_0H}
  \\
  \;-\;
  \frac{\im\gamma^2}{(2\,\im+k_0H)^2}
  \,\cdot\Bigg\{
  \frac{2}{3}\;-\;
  \frac{2\im+k_0H}{k_0H}
  \bigg\}
  \,(k_0 H)(k_\para H)^2.
\end{multline*}
It is of interest to note that the above approximate relations entail
\begin{equation*}
  R_{\text{ENZ}}+e^{\im k_0 H}T_{\text{ENZ}}\simeq 1\quad \mbox{and}\quad
  |R_{\text{ENZ}}|^2+|T_{\text{ENZ}}|^2\simeq 1.
\end{equation*}

%%%%%%%%%%%%%%%%%%%%%%%%%%%%%%%%%%%%%%%%%%%%%%%%%%%%%%%%%%%%%%%%%%%%%%%%%%%%%%%%
%%%%%%%%%%%%%%%%%%%%%%%%%%%%%%%%%%%%%%%%%%%%%%%%%%%%%%%%%%%%%%%%%%%%%%%%%%%%%%%%
%%%%%%%%%%%%%%%%%%%%%%%%%%%%%%%%%%%%%%%%%%%%%%%%%%%%%%%%%%%%%%%%%%%%%%%%%%%%%%%%
\section{Homogenization theory, corrector field and resonances}
\label{sec:homogen-reson}

In this section, we revisit the established general theory of
homogenization for periodic layered plasmonic
structures~\cite{Maier18,Maier19b}. In this theory, the corrector field can
encode subwavelength plasmonic resonances inherent to the geometry of the
constituent 2D material. Although the general homogenization theory has
already been derived in \cite{Maier19b} for arbitrary geometries of 2D
materials in periodic structures (Appendix~\ref{app:hom-gen-th}), it is now
tailored to the computation of experimentally observable quantities for
practically appealing configurations, e.g., layered structures with
graphene nanoribbons.

We use a simplified set of material parameters for the dielectric host and
conducting sheet. In Sec.~\ref{subsec:hom-Fresnel-diagonal_e}, we derive
formulas for the homogenized Fresnel coefficients, by adopting ingredients
of the general homogenization theory in the present case with structures of
finite thickness. To avoid technical complications, we restrict our
attention to models that yield a \emph{diagonal effective} permittivity
matrix under TM polarization (cf. Sec.~\ref{sec:analysis}). The resulting
formulas for the optical coefficients provide a nontrivial extension to
their counterparts for infinite, translation invariant sheets; cf.
Eq.~\eqref{eq:homogenized_layers}. The role of the corrector field is
discussed in Sec.~\ref{subsec:corr-phys}.

More precisely, our procedure regarding the Fresnel coefficients consists
of the following stages.
First, we provide the effective permittivity tensor, $\teff$, that results
from the homogenization of a periodic array of conducting sheets with
arbitrary geometry. For general configurations, this $\teff$ has been
derived from a two-scale asymptotic expansion for the fields obeying
Maxwell's equations in the limit $k_0 d\to 0$~\cite{Maier19b};
alternatively, one may use the Bloch wave theory for simple enough
geometries~\cite{Maier18}. The methodology of asymptotic expansions has the
advantage that it is not limited to plane wave solutions.
Second, we replace the layered plasmonic crystal (Fig.~\ref{fig:sandwich})
by a single \emph{homogenized} anisotropic dielectric slab that has the
permittivity tensor $\teff$ and the same thickness as the layered
structure. We then compute directly the corresponding, effective Fresnel
coefficients $R^{\text{eff}}$ and $T^{\text{eff}}$ when $\teff$ is diagonal
(Sec.~\ref{subsec:hom-Fresnel-diagonal_e}). The results of this computation
are relevant to nanoribbons (Sec.~\ref{sec:computation}).

To simplify the exposition, we assume that $\e_0=\e_{\text{s}}$. To avoid
complications due to the microscale material parameters per se,  we posit
that the dielectric host has the (isotropic and homogeneous) permittivity
$\te=\text{diag}(\varepsilon,\varepsilon,\varepsilon)$ and the conducting
sheet has the scalar surface conductivity $\sigma$. The surface of the
sheet is assumed to be smooth enough, having a uniquely defined normal
vector at every point of its interior (away from the boundary). The sheet
can have edges. We alert the reader that the generality of our
homogenization result for $\teff$ mainly concerns the sheet geometry (see
Appendix~\ref{app:hom-gen-th}).
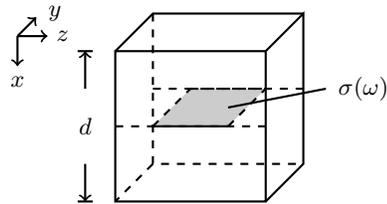
\begin{figure}
 \centering
 \begin{tikzpicture}[scale=1.0]
    \path [thick, ->, draw] (-2.3, 1.2) -- (-1.9, 1.2);
    \path [thick, ->, draw] (-2.3, 1.2) -- (-2.3, 0.8);
    \path [thick, ->, draw] (-2.3, 1.2) -- (-2.05, 1.45);
    \node at (-2.3, 0.6) {$x$};
    \node at (-1.7, 1.2) {$z$};
    \node at (-1.8, 1.5) {$y$};
    \path [thick, draw] (-1.5,1) -- (-1.3,1);
    \path [thick, draw] (-1.5,-1) -- (-1.3,-1);
    \path [thick, ->, draw] (-1.4,0.5) -- (-1.4,1);
    \node at (-1.4,0) {$d$};
    \path [thick, ->, draw] (-1.4,-0.5) -- (-1.4,-1);
    \path [thick, draw]
      (-1, -1) -- (1, -1) -- (1, 1) -- (-1, 1) -- cycle;
    \path [thick, draw] (1,-1) -- (1.5, -0.5) -- (1.5, 1.5) -- (1, 1);
    \path [thick, draw] (-1,1) -- (-0.5, 1.5) -- (1.5, 1.5);
    \path [thick, draw, dashed]
      (-0.5, 1.5) -- (-0.5, -0.5) -- (1.5, -0.5);
    \path [thick, draw, dashed] (-0.5, -0.5) -- (-1.0, -1.0);
    \path [fill=black, fill opacity=0.2]
      (-0.5, 0) -- (0.5, 0) -- (1.0, 0.5) -- (-0.0, 0.5) -- (-0.5, 0.0);
    \path [thick, draw] (-0.5, 0) -- (0.5, 0);
    \path [thick, draw, dashed] (0.5, 0) -- (1.0, 0.5);
    \path [thick, draw, dashed] (-0.0, 0.5) -- (1.0, 0.5);
    \path [thick, draw, dashed] (-0.5, 0.0) -- (-0.0, 0.5);
    \path [thick, draw, dashed] (-1, 0) -- (1, 0);
    \path [thick, draw, dashed] (-0.5, 0.5) -- (1.5, 0.5);
    \path [thick, draw] (0.5,0.25) -- (1.8,0.5);
    \node at (2.3,0.5) {$\sigma(\omega)$};
  \end{tikzpicture}
  \caption{Schematic of representative volume element (box) in nanoribbon
    configuration: The conducting strip lies in the $yz$-plane, with each
    edge being parallel to the $y$-axis. For the purpose of periodic
    homogenization, the representative volume element is repeated
    periodically with interlayer spacing equal to $d$ (in the
    $x$-direction).}
  \label{fig:nanoribbons-rve}
\end{figure}
It can be shown that for a representative volume element of linear size
equal to $d$ in all directions (Fig.~\ref{fig:nanoribbons-rve}), the matrix
elements of the effective permittivity $\teff$ take the form
\begin{align}
  \eff_{ij} \;=\; \varepsilon\delta_{ij} -\frac{\sigma(\omega)}{\im\omega
  d^3}\int_{\Sigma}
  \left[\boldsymbol\tau_j+\nabla_{\boldsymbol\tau}\chi_j(\vr)\right]
  \cdot\boldsymbol e_i\,\db\vr.
  \label{eq:eff-general}
\end{align}
Here, $\chi_i$ is the $i$-th component of the corrector field $\boldsymbol
\chi$ (further described below); $\Sigma$ is the surface region of the
conducting sheet inside the representative volume element; $\vr=(x,y,z)$
and $\delta_{ij}$ is Kronecker's delta with $i,\,j=x,\,y,\,z$. To define
the remaining quantities in Eq.~\eqref{eq:eff-general}, let $\boldsymbol
\nu$ denote the (uniquely defined) unit vector normal to $\Sigma$.
Accordingly, $\boldsymbol\tau_i(\vr)$ is the projection of the $i$-directed
Cartesian vector, $\boldsymbol e_i$, to the plane tangential to $\Sigma$ at
point $\vr$, viz., $\boldsymbol \tau_i=\boldsymbol e_i-(\boldsymbol
e_i\cdot\boldsymbol\nu)\boldsymbol \nu$; and $\nabla_{\vec\tau}$ is the
surface gradient on $\Sigma$, i.e., $\nabla_{\vec\tau}=\nabla -(\nabla\cdot
\boldsymbol \nu)\boldsymbol \nu$. Note that $\Sigma$ is finite. In the
special case where $\Sigma$ is flat and lies on the $yz$-plane, we have
$\boldsymbol\tau_j\cdot\boldsymbol e_i=\delta_{ij}$ if $i,\,j=y,z$ and
$\boldsymbol\tau_j\cdot\boldsymbol e_i=0$ otherwise.
Equation~\eqref{eq:eff-general} describes a suitable weighted
\emph{average} over microscale ($d$-dependent) details. The weight is
determined by the corrector field $\vec\chi$ which is determined by a
Helmholtz-like boundary-value problem in the representative volume element
(see Appendix~\ref{app:hom-gen-th}). This field captures fine-scale,
low-order lateral plasmonic resonances that are possibly excited in the 2D
material (see Sec.~\ref{subsec:comput-nanorb}).

%%%%%%%%%%%%%%%%%%%%%%%%%%%%%%%%%%%%%%%%%%%%%%%%%%%%%%%%%%%%%%%%%%%%%%%%%%%%%%%%

\subsection{Homogenized Fresnel coefficients in TM polarization}
\label{subsec:hom-Fresnel-diagonal_e}

Next, we use the general homogenization result~\eqref{eq:eff-general} for
$\teff$ in order to compute the effective Fresnel coefficients of a layered
plasmonic structure in the limit $k_0 d\to 0$ with finite total thickness
(see Fig.~\ref{fig:sandwich}). The incident plane wave is TM polarized. For
the sake of simplicity, we focus on geometries for which the matrix $\teff$
is diagonal, $\teff=\text{diag}(\eff_x,\eff_y,\eff_z)$, where possibly
$\eff_x\neq \eff_y\neq \eff_z\neq \eff_x$. This case accounts for
configurations with infinitely extended sheets, nanoribbons and circular
nanotubes~\cite{Maier19b}. Note that this type of anisotropic permittivity
$\teff$ may result even from the homogenization of isotropic dielectric
hosts with conducting sheets. Numerical simulations based on our
homogenization results for nanoribbons are presented in
Sec.~\ref{sec:computation}.

The goal in this section is to replace the multilayer system of
Fig.~\ref{fig:sandwich} by a single-layer, continuous medium with
dielectric permittivity equal to $\teff$. This medium is of course located
between the air and dielectric substrate and has thickness equal to $H$. To
calculate the coefficients $R$ and $T$ for TM polarization, we apply the
transfer matrix approach of Sec.~\ref{subsec:eval-Fresnel} with $N=1$,
$\e_\perp=\eff_x$ and $\e_\para=\eff_z$; thus, we replace $d$ by $H$.
Because of our assumption for a TM-polarized incident plane wave, the
matrix element $\eff_y$ does not enter the calculation.

By inspection of the transfer matrix procedure, we realize that we can
invoke Eq.~\eqref{eq:fresnel} with $t_{11}=t_{22}=1$ and $t_{12}=t_{21}=0$
(see Sec.~\ref{subsec:eval-Fresnel}). Consequently, we find
\begin{align}
\label{eq:homogenized_general}
  \begin{cases}
    \begin{aligned}
      R^{\text{eff}}\;&=\;-\;\frac%
      {\eb_{\text{s}+}\,\eb_{0-} -\eb_{\text{s}-}\,\eb_{0+}\,e^{\im 2 \beff H}}
      {\eb_{\text{s}+}\,\eb_{0+}-\eb_{\text{s}-}\,\eb_{0-}\,e^{\im 2\beff H}},
      \\
      T^{\text{eff}}\;&=\;e^{-\im k_{x,\text{s}}H}\,
      \frac{(\eb_{0+}^2-\eb_{0-}^2)e^{\im \beff H}}
      {\eb_{\text{s},+}\,\eb_{0+}-\eb_{\text{s}-}\,\eb_{0-}\,e^{\im2\beff H}}
    \end{aligned}
  \end{cases}
\end{align}
where $\eb_{\ell\pm}$ ($\ell=0,\,\text{s}$) are defined by
\begin{align*}
  \eb_{\ell\pm}
  \;=\;1\,\pm\,\frac{\e_{\ell}}{\eff_\perp}
  \frac{(\keff_\perp)^2-k_z^2}{\beff\,k_{x,\ell}};\quad
  \keff_\perp=\omega\sqrt{\mu\eff_\perp}
\end{align*}
with $\eff_\perp=\eff_x $. We invoke the effective wavenumber
\begin{align}
  \label{eq:beff}
  \beff =
  \sqrt{\frac{\eff_\para}{\eff_\perp}\big[(\keff_\perp)^2-k_z^2\big]}
  \end{align}
where $\eff_\para=\eff_z$.

We should point out that Eq.~\eqref{eq:homogenized_general} is valid for a
wide class of sheet geometries, subject to the diagonal character of
$\teff$, in contrast to Eq.~\eqref{eq:homogenized_layers}. The main
difference of the present model for $\teff$ from its counterpart  of the
transfer matrix approach  (Sec.~\ref{sec:analysis}) is the effect of the
corrector field, $\boldsymbol\chi$. The derivation of the effective Fresnel
coefficients for a more general, homogeneous but non-diagonal $\teff$ is
tractable but non-essential for our scope.

It is worthwhile to check that the coefficients of
Eq.~\eqref{eq:homogenized_general} correctly reduce to the corresponding
Fresnel coefficients for the fully translation invariant sheets of
Sec.~\ref{sec:analysis}. In this case, the boundary value problem for
$\boldsymbol \chi$ (Appendix~\ref{app:hom-gen-th}) yields $\boldsymbol
\chi=0$. By Eq.~\eqref{eq:eff-general}, we obtain $\eff_x = \e$ and $\eff_z
= \e + \im\sigma/(\omega d)=\eff_y=\eff_\para$. For normal incidence of the
incoming plane wave ($k_z=0$), we have
\begin{align*}
  \beff = \sqrt{\omega^2\mu\eff_\para} =
  \sqrt{\beta^2+\frac{\im\omega\mu\sigma}{d}}
\end{align*}
where $\beta^2=\omega^2\mu\e=k_\para^2$, by the notation of
Eq.~\eqref{eq:kx-beta-disp}. The substitution of the above value for the
effective wavenumber $\beff$ into Eq.~\eqref{eq:homogenized_general} yields
the homogenized Fresnel coefficients in agreement with
Eq.~\eqref{eq:homogenized_layers} if  $\e_0=\e_{\text{s}}$.

%%%%%%%%%%%%%%%%%%%%%%%%%%%%%%%%%%%%%%%%%%%%%%%%%%%%%%%%%%%%%%%%%%%%%%%%%%%%%%%%
\subsection{Corrector field in nanoribbon geometry}
\label{subsec:corr-phys}

We now illustrate the role of the corrector field $\boldsymbol\chi$ by
choosing to focus on the nanoribbon configuration with an isotropic
dielectric host (Fig.~\ref{fig:nanoribbons-rve}). In this geometry, as
outlined in the context of the general homogenization theory
(Appendix~\ref{app:hom-gen-th}), all components of $\boldsymbol\chi$ vanish
identically except $\chi_z$. Thus, by Eq.~\eqref{eq:eff-general} the
effective permittivity tensor is written as
$\teff=\text{diag}(\eff_x,\eff_y,\eff_z)$ where
\begin{subequations}\label{eqs:hom-eta}
\begin{align}
  \eff_x \;&=\; \e,
  \quad \eff_y=\e-\eta(\omega),\\
  \eff_z \;&=\;
  \e-\eta(\omega)\,
  \frac{1}{d^2}\int_\Sigma
  \big[1+\partial_z\chi_\para(\vr)\big]\,\db\vr.
\end{align}
\end{subequations}
In the above, $\eta(\omega) = \frac{\sigma(\omega)}{\im\omega d}$, which is
further discussed in Sec.~\ref{sec:computation}; and the corrector
$\chi_\para(\vr)=\chi_z(\vr)$ solves the associated cell problem
(Appendix~\ref{app:hom-gen-th}). More generally, if the dielectric host has
permittivity tensor $\te=\text{diag}(\e_\perp,\e_\para,\e_\para)$ with
$\e_\perp\neq \e_\para$ then $\e\to \e_\perp$ in $\eff_x$ whereas $\e\to
\e_\para$ in both $\eff_y$ and $\eff_z$. Let us recall that, for a
TM-polarized incident plane wave, the Fresnel coefficients $R^{\text{eff}}$
and $T^{\text{eff}}$ are only affected by the parameters $\eff_x$ and
$\eff_z$. We should add the remark that Eqs.~\eqref{eqs:hom-eta} are
applicable to the configuration with infinite, translation invariant sheets
for $\chi_\para=0$, if the media parameters are kept fixed.

The above corrector field, $\chi_\para(\vr)$, in principle encodes the
response of the plasmonic microstructure to all possible surface
excitations by local plane waves~\cite{Maier19b}. In particular, this
response may include short-scale, subwavelength surface modes on the 2D
material, which we interpret as \emph{(lateral) SPP
resonances}~\cite{MaierML18}. These modes are excited on the 2D material
because of the ribbon edges, and its finite width, as the asymptotically
``slow'' macroscopic electromagnetic wave solution approaches a plane wave
in the homogenization limit~\cite{Maier19b}.

This interpretation is consistent with the following observation. In the
special geometry with infinite, translation invariant sheets
(Sec.~\ref{sec:analysis}), the conducting material does not admit such a
surface excitation, and the corrector $\boldsymbol\chi(\vr)$ vanishes
identically~\cite{Maier19b}. In principle, the absence of microscale
surface excitations on the 2D material should be equivalent to a vanishing
corrector for the cell problem; consequently,
$\eff_y=\eff_z=\e_\para-\eta(\omega)\frac{1}{d^2}
\int_\Sigma\text{d}\boldsymbol r$. In contrast, the
configuration with nanoribbons shows a dominant influence of the corrector
field $\chi_z(\vr)$ when the frequency $\omega$ is close to resonance
frequencies; see Sec.~\ref{sec:computation} for numerical results.

%%%%%%%%%%%%%%%%%%%%%%%%%%%%%%%%%%%%%%%%%%%%%%%%%%%%%%%%%%%%%%%%%%%%%%%%%%%%%%%%
%%%%%%%%%%%%%%%%%%%%%%%%%%%%%%%%%%%%%%%%%%%%%%%%%%%%%%%%%%%%%%%%%%%%%%%%%%%%%%%m
%%%%%%%%%%%%%%%%%%%%%%%%%%%%%%%%%%%%%%%%%%%%%%%%%%%%%%%%%%%%%%%%%%%%%%%%%%%%%%%m

\section{Computational results}
\label{sec:computation}

In this section, we numerically compare a quantity, the complementary
transmission spectrum (defined below), of a fully \emph{layered} structure
to the respective result of the homogenization procedure. Our goal is to
quantitatively assess the accuracy of homogenized models for the
computation of wave transmission through plasmonic crystals for frequencies
and geometries of possible practical interest. Of particular significance
in applications is the dependence of the optical coefficients on the number
of layers~\cite{Yanetal2012}, which we study in some detail below. We
assume that the electromagnetic field has TM polarization.

We choose to focus on two geometries with conducting isotropic 2D
materials. One configuration consists of infinite, mutually parallel sheets
(Fig.~\ref{fig:sandwich}); and another consists of nanoribbons
(Fig.~\ref{fig:nanoribbons-rve}). The former setting serves as a `reference
case', since it allows us to apply the exact results of the transfer matrix
approach from Sec.~\ref{sec:analysis}. In this geometry, however, there are
\emph{no microscale lateral resonances}. In contrast, the nanoribbon
configuration enables the appearance of such resonances; in the
homogenization limit, these effects can be captured by the corrector field,
as we show below. For the nanoribbon case, we solve the full Maxwell system
for the electromagnetic field via the finite element
method~\cite{Maier17,MaierML18}; and compare the result to the respective
homogenization outcome through the numerical solution of the boundary value
problem for the corrector field (see Sec.~\ref{sec:homogen-reson}). In both
cases of layered configurations, we assume that the 2D material is doped
monolayer graphene. The microstructure includes an isotropic dielectric
host ($\e_\para=\e_\perp=\e$). The macroscopic structure of thickness $H$
lies between vacuum and a dielectric substrate (with permittivity
$\e_{\text{s}}>\e_0$). All media are nonmagnetic ($\mu=\mu_0$ for
definiteness).

%%%%%%%%%%%%%%%%%%%%%%%%%%%%%%%%%%%%%%%%%%%%%%%%%%%%%%%%%%%%%%%%%%%%%%%%%%%%%%%%
\subsection{Preliminaries}
\label{subsec:comput_prelim}

First, we outline the setup of our numerical computations. For the
homogenized Fresnel coefficients, we make use of
Eqs.~\eqref{eq:homogenized_general}, which in principle incorporate the
effect of the vector valued corrector field $\boldsymbol \chi$, in
conjunction with Eqs.~\eqref{eqs:hom-eta}. In the reference case (infinite
planar sheets), the corrector field vanishes identically ($\boldsymbol
\chi=0$). Recall that ingredients of the homogenization theory for the
nanoribbon geometry are spelled out in Sec.~\ref{subsec:corr-phys}.

We apply a non-dimensionalization of the relevant equations. In particular,
we use the following rescaling of key parameters:
\begin{align*}
  \tilde\omega = \frac{\hbar\omega}{E_F},
  \qquad
  \tilde k = \frac{k}{k_0},
  \qquad
  \tilde\sigma(\tilde\omega) = \sqrt{\frac{\mu_0}{\e_0}}\,\sigma(\omega),
\end{align*}
where $E_{\text{F}}$ denotes the Fermi energy, $k$ is any relevant
wavenumber (e.g., $k_{\text{s}}$, $k_\perp$ and $\beff$), $\hbar$ is the
reduced Planck constant, and $\mu_0=\mu$. We combine the above rescaling
with the Drude model for the scalar surface conductivity of the (isotropic)
2D material~\cite{Maier17}. Hence, we use the following dimensionless
surface conductivity~\cite{Maier17}:
\begin{align*}
  \tilde\sigma_{\text{Drude}}(\tilde\omega) = \frac{i\,\tilde\omega_{\text{p}}}
  {\tilde\omega + i/\tilde\tau},
  \quad
  \tilde\omega_{\text{p}} =
  \frac{4\,e^2}{4\pi\varepsilon_0\hbar c_0}
  =4\,\alpha.
\end{align*}
Here, $\tilde\tau=(E_{\text{F}}/\hbar)\tau$, $\tau$ is the
(phenomenological) relaxation time of the Drude model, and $\alpha$ denotes
the fine structure constant; as usual, $e$ stands for the elementary
(electron) charge and $c_0$ is the speed of light in vacuum.

In our numerics, we choose to compute the quantity
\begin{align}\label{eq:complem-transmission}
  \FTC = 1 - \big|T(\tilde\omega)\big|^2,
\end{align}
which we refer to as the \emph{the complementary transmission spectrum} for
the layered plasmonic structure of interest. This $\FTC$ is akin
(but not identical) to the extinction spectrum of layered structures, which
is usually measured in experiments~\cite{Yanetal2012}. An advantage of
using this $\FTC(\tilde \omega)$ is that it satisfies the inequality $0<
\FTC< 1$. We compute $\FTC$ by: (i) the transfer matrix approach as well as
the explicit homogenization formula for $T$ (with zero corrector field) for
the reference case (Sec.~\ref{subsec:comput-planar}); and (ii) the
numerical solution of the Maxwell system as well as the homogenization
procedure with a corrector field for the nanoribbon configuration
(Sec.~\ref{subsec:comput-nanorb}).

We pay particular attention to the \emph{deviation} of the homogenized
version of $\FTC$ from the corresponding quantity for the
original layered structure. We refer to this deviation as the
\emph{homogenization error}. Recall that, in the homogenized problem, the
layered structure is replaced by a slab of equal total thickness with an
effective continuous medium. We extend the definition of the homogenization
error to the computation of resonance frequencies in the nanoribbon
configuration (Sec.~\ref{subsec:comput-nanorb}).

To illustrate the homogenization error computationally, we choose and fix
the relevant material parameters as follows. The dielectric permittivity of
the substrate is $\e_{\text{s}} = 4.4\e_0$, while the medium above the
layered structure is vacuum (with permittivity $\e_0$). The value for
$\e_{\text{s}}$ used here is typical for quartz~\cite{Davies2018}.  We
assume that the dielectric host is isotropic with $\e_\para =
\e_\perp=2.3\e_0$. This choice roughly corresponds to the permittivity
values for polymer-based buffer materials in layered plasmonic structures,
if one ignores the frequency dependence of the permittivity tensor
$\te$~\cite{Chang17}.

By Eqs.~\eqref{eqs:hom-eta} of the homogenization procedure, our parameter
rescaling leads to the following weight for elements of the effective
permittivity tensor~\cite{Maier19b}:
\begin{align*}
  \tilde\eta(\tilde\omega)\;=\;
  \frac{\tilde\omega_p}{\tilde\omega^2(\tilde\omega+\im/\tilde\tau)\,\bar
  d},
\end{align*}
where $\bar d=\tilde d/\tilde \omega$ and $\tilde d=k_0 d$; thus, $\bar d$
is frequency independent. In our numerical computations, we use the
(dimensional) relaxation time $\tau=0.4\,$ps at the Fermi energy
$E_{\text{F}}=0.4\,$eV, which are typical for monolayer graphene. These
choices imply the (non-dimensional) parameter value $\tilde\tau = 243.2$.
We also fix the interlayer spacing to $d=25\,$nm, which yields the
parameter value $\bar d = 0.05068$. This choice of spacing is compatible
with experiments on wave transmission through stacks consisting of graphene
sheets and insulator slabs~\cite{Yanetal2012}.

%%%%%%%%%%%%%%%%%%%%%%%%%%%%%%%%%%%%%%%%%%%%%%%%%%%%%%%%%%%%%%%%%%%%%%%%%%%%%%%%
\subsection{Infinite planar conducting sheets}
\label{subsec:comput-planar}

Next, we carry out computations for the layered configuration of the
reference case (Fig.~\ref{fig:sandwich}). Our goal is to assess the
accuracy of the homogenization results. In particular, we point out the
negligible homogenization error even for small values of the number $N$ of
layers. We remind the reader that the (scaled) interlayer spacing $\bar d$
is kept fixed in our computations.

As a starting point, in Fig.~\ref{fig:beff} we show plots of the
($N$-independent) real and imaginary parts of the non-dimensional effective
wavenumber, $\tilde{\beta}^{\text{eff}}=\beta^{\text{eff}}/k_0$, as a
function of the rescaled frequency, $\tilde\omega$. The parameter
$\tilde{\beta}^{\text{eff}}$ is computed by Eq.~\eqref{eq:beff} with the
appropriate, homogenized tensor permittivity.

\begin{figure}
  \centering
  \input{beta-eff.tex}
  \vspace{-1.5em}
  \caption{\footnotesize%
    (Color online) Real and imaginary parts of rescaled, non-dimensional
    effective wavenumber , $\tilde{\beta}^{\text{eff}}$, as a function of
    $\tilde\omega$ by Eq.~\eqref{eq:beff}, for the geometry with infinite
    planar sheets.}
  \label{fig:beff}
  \vspace{1.0em}

  \input{extinction.tex}
  \vspace{-1.5em}
  \caption{\footnotesize%
    (Color online) Complementary transmission spectrum $\FTC$
    for infinite planar sheets. The computations are carried out via:
    transfer matrix approach (solid line); and homogenization procedure
    (dashed line). We use $N=4,\,8,\,16$ layers in each approach. The
    vertical dashed line indicates the ENZ condition.}
  \label{fig:extinction}
  \vspace{1.0em}

  \input{error.tex}
  \vspace{-1.5em}
  \caption{\footnotesize%
    (Color online) Relative homogenization error for complementary
    transmission spectrum $\FTC$ for the geometry with infinite
    planar sheets. We use $N=4,\,8,\,16$ layers.}
  \label{fig:error}
\end{figure}

In Fig.~\ref{fig:extinction}, we plot the complementary transmission
spectrum $\FTC$ within the (exact) transfer matrix approach as
well as the (approximate) homogenization procedure. We consider  $N=4,\,8$
and $16$ layers. In these plots, the rescaled total  thickness of the
plasmonic structure is $\bar H = \bar d N=k_0 H$ where $H=Nd$ takes the
values $100\,$nm, $200\,$nm and $400\,$nm, respectively, as $N$ varies. In
Fig.~\ref{fig:error}, we show the relative homogenization error for the
complementary transmission spectrum in a wide range of the (rescaled)
frequency $\tilde\omega$.

A few remarks on the displayed numerical results are in order. Regarding
Fig.~\ref{fig:beff}, the ENZ condition is attained at the ($N$-independent)
frequency $\tilde\omega\approx 0.63$. The computation of $T^{\text{eff}}$
at the ENZ condition (Sec.~\ref{sec:enz}) gives a value for this
coefficient in excellent agreement with the corresponding value shown in
Fig.~\ref{fig:extinction} (dashed curve). By inspection of
Fig.~\ref{fig:error}, we see that the relative homogenization error remains
well below $1\,\%$ for a wide frequency range. A maximum relative error of
about $10\,\%$ occurs at $\tilde\omega\approx0.2,0.3,0.46$ when
$N=4,\,8,\,16$, respectively. This error is seen to decrease almost inverse
linearly with the number, $N$, of layers if $N$ is sufficiently large.
The local minimum in the complementary transmission spectrum for $N=16$
(Fig.~\ref{fig:extinction}) can be attributed to an \emph{interlayer
resonance} (in the $x$-direction), to be distinguished from the lateral
resonances of the nanoribbon geometry (Sec.~\ref{subsec:comput-nanorb}).
For increasing total thickness $H$, which scales linearly with $N$ (for
fixed spacing $d$), these resonances are shifted to lower frequencies.
Indeed, by numerically computing $\mathfrak{T}_{\text{c}}$ for the cases
with $N=32, 64, 128$, which are not displayed in the present plots, we
observe roughly a doubling in the number of the above minima in the
frequency range  $0\le \tilde\omega\le 2$ every time $N$ is doubled. We
expect this trend to persist for higher values of $N$.

%%%%%%%%%%%%%%%%%%%%%%%%%%%%%%%%%%%%%%%%%%%%%%%%%%%%%%%%%%%%%%%%%%%%%%%%%%%%%%%%
\subsection{Nanoribbon geometry}
\label{subsec:comput-nanorb}

In this section, we study numerically the wave transmission through the
layered structure that contains mutually parallel nanoribbons.
In our numerical simulations, we set the width of each nanoribbon as well
as the lateral spacing between nanoribbons in the $yz$-plane equal to $d$.
Hence the representative volume element has a linear size equal to $2d$ in
the $z$-direction but the interlayer spacing is kept equal to $d$; cf.
Fig.~\ref{fig:nanoribbons-rve}.

We compare outcomes of our numerics from two main approaches: One approach
is the direct numerical solution of the full Maxwell system via the finite
element method (for transmission through the layered
configuration)~\cite{Maier17}; and another is the homogenization procedure
(Sec.~\ref{subsec:corr-phys}). For the homogenized structure, the boundary
value problem for the corrector field $\boldsymbol\chi$ is solved
numerically by the finite element method~\cite{Maier19b}. In this setting,
we describe (lateral) SPP-related resonances inherent to each nanoribbon.

In Fig.~\ref{fig:ribbons-beff}, we plot the ($N$-independent) real and
imaginary parts of the rescaled effective wavenumber,
$\tilde{\beta}^{\text{eff}}$, as a function of the rescaled frequency,
$\tilde\omega$. The parameter $\tilde{\beta}^{\text{eff}}$ is computed by
Eq.~\eqref{eq:beff} with the effective tensor permittivity of the
nanoribbon geometry. In the present case, this computation involves a
nontrivial corrector field.

The complementary transmission spectrum, $\FTC$, computed by
both the direct and homogenization approaches is shown in
Fig.~\ref{fig:ribbons-extinction}. The numerical computations here are
carried out for $N=4$ and $8$ layers. In Fig.~\ref{fig:ribbons-error}, we
display the relative homogenization error versus frequency $\tilde\omega$,
in regard to the computation of $\FTC$ for each of the chosen
values for $N$.

\begin{figure}
  \centering
  \input{ribbons-beta-eff.tex}
  \vspace{-1.5em}
  \caption{\footnotesize%
    (Color online) Real and imaginary parts of rescaled, non-dimensional
    effective wavenumber, $\tilde{\beta}^{\text{eff}}$, as a function of
    non-dimensional frequency $\tilde\omega$ by Eq.~\eqref{eq:beff}, for
    the nanoribbon geometry.}
  \label{fig:ribbons-beff}
  \vspace{1.0em}

  \input{ribbons-extinction.tex}
  \vspace{-1.5em}
  \caption{\footnotesize%
    (Color online) Complementary transmission spectrum $\FTC$
    for the nanoribbon geometry. The computations are carried out via:
    numerical solution of transmission problem for fully layered structure
    (solid line); and homogenization procedure (dashed line). We use
    $N=4,\,8$ layers in each approach.}
 \label{fig:ribbons-extinction}
  \vspace{1.0em}

  \input{ribbons-error.tex}
  \vspace{-1.5em}
  \caption{\footnotesize%
    (Color online) Relative homogenization error for complementary
    transmission spectrum $\FTC$ for nanoribbon structure. We
    use $N=4,\,8$ layers.}
  \label{fig:ribbons-error}
\end{figure}

In our numerics, the lateral resonances manifest in the form of local peaks
of the complementary transmission spectrum at certain frequencies, for
fixed number, $N$, of layers; see Fig.~\ref{fig:ribbons-extinction}. We
reiterate that such resonances do \emph{not} occur in the configuration
with infinite conducting sheets (Sec.~\ref{subsec:comput-planar}).

It is worthwhile to further quantify these lateral resonances. Let
$\tilde\omega_n$ denote the relevant (non-dimensional) resonance
frequencies of the complementary transmission spectrum computed directly
for the actual layered structure; $n$ is a positive integer counting these
frequencies in ascending order ($\tilde\omega_{n+1}>\tilde\omega_{n}$ with
$n=1,\,2,\,\ldots$). Here, we set $n$ equal to $1$ for the lowest resonance
frequency computed in our numerics. In
Table~\ref{tab:resonant_frequencies}, we list the first eight
($n=1,\,2,\,\ldots,\,8$) of these resonance frequencies, when the number of
layers is $N=4$ and $N=8$.

We also compute the corresponding resonance frequencies,
$\tilde{\omega}^{\text{eff}}$, from the homogenization approach. For this
purpose, we locally fit a Lorentzian to the frequency response of the
effective wavenumber $\tilde{\beta}^{\text{eff}}(\tilde\omega)$ near each
resonance; see Fig.~\ref{fig:ribbons-beff}. We report the results in
Table~\ref{tab:resonant_frequencies}.

We also provide the  resonance frequencies, $\tilde\omega_{n}^{\text{D}}$,
that come from the ($N$-independent) ``Dirichlet approximation'' pertaining
to a single, \emph{isolated} nanoribbon. This approximation is derived from
the SPP excitation along a single conducting strip of finite width as
follows (see Fig.~\ref{fig:nanoribbons-rve}): At the edges of the
nanoribbon, impose homogeneous (zero) Dirichlet conditions to the generated
unperturbed SPP, a suitable trigonometric function of $z$, for the
$z$-component of the electric field~\cite{MaierML18}. Accordingly, we find
that the Dirichlet approximation $\tilde\omega_{n}^{\text{D}}$ is obtained
by the relation
\begin{align}
  \label{eq:dirichlet_approximation}
  \frac{\tilde w(\tilde\omega_{n}^{\text{D}})\,\tilde
  k_{\text{spp}}(\tilde\omega_{n}^{\text{D}})}{2\,\pi}
  \;=\; \frac{2n - 1}{2}\quad (n=1,\,2,\,\ldots),
\end{align}
where $\tilde w(\tilde\omega_{n}^{\text{D}}) =
\tilde\omega_{n}^{\text{D}}\bar d$ is the width of an individual nanoribbon
and
\begin{align*}
  \tilde k_{\text{SPP}}(\tilde\omega_{n}^{\text{D}})\;=\;
  \sqrt{\tilde\e_\para -
  \frac{4\,\tilde\e_\para^2}{\tilde\sigma(\tilde\omega_{n}^{\text{D}})^2}}
\end{align*}
denotes the rescaled SPP wavenumber~\cite{MaierML18} at frequency
$\tilde\omega_{n}^{\text{D}}$. Here, $\tilde\e_\para = \e_\para/\e_0$ is
the relative permittivity.

We close this subsection with a few more remarks on our numerics. An
inspection of Fig.~\ref{fig:ribbons-error} indicates that the relative
homogenization error (versus $\tilde\omega$) does not exceed about $1\,\%$
for a wide frequency range. In fact, this deviation is exacerbated, with
the relative error having local maxima in $\tilde\omega$ that may reach
about $10\,\%$, near the resonance frequencies $\tilde\omega_n$. In regard
to the calculation of the frequencies $\tilde\omega_n$, overall we observe
a very good agreement between the directly computed values of these
frequencies (for the actual layered structure) and their counterparts,
$\tilde\omega^{\text{eff}}_{n}$, for the homogenized problem. More
precisely, the related deviation for $N=8$ does not exceed $2.1\%$  (see
Table~\ref{tab:resonant_frequencies}). A close inspection of
Fig.~\ref{fig:ribbons-extinction} reveals that the behavior of
$\FTC$ near each resonance can be described (locally in
$\tilde\omega$) by a Lorentzian, as expected~\cite{Maier19b}. The width of
each Lorentzian depends on the dissipation, i.e., the real part of the
surface conductivity $\sigma(\omega)$. According to the Drude model
$\tilde\sigma_{\text{Drude}}(\tilde\omega)$ used here, this dissipation is
controlled by the (rescaled) relaxation time $\tilde\tau$.

\begin{table}
  \centering
  \begin{tabular}{lccccc}
    \toprule
    & $n$ & $\tilde\omega^{\text{eff}}_{n}$ & $\tilde\omega_{n}^{\text{D}}$ &
    $\tilde\omega_{n}\,(N=4)$ & $\tilde\omega_{n}\,(N=8)$ \\[0.1em]
    \cmidrule(lr){2-2}
    \cmidrule(lr){3-3}
    \cmidrule(lr){4-4}
    \cmidrule(lr){5-5}
    \cmidrule(lr){6-6}
     & 1 & 0.6501 & 0.7327 ($13\,\%$)& 0.6446 ($0.8\%$)& 0.6484 ($0.3\%$)\\
     & 2 & 1.0237 & 1.0568 ($3.2\%$)&  1.0213 ($0.2\%$)& 1.0215 ($0.2\%$)\\
     & 3 & 1.2287 & 1.2530 ($2.0\%$)&  1.2241 ($0.4\%$)& 1.2242 ($0.4\%$)\\
     & 4 & 1.3814 & 1.4017 ($1.5\%$)&  1.3735 ($0.6\%$)& 1.3733 ($0.6\%$)\\
     & 5 & 1.5062 & 1.5242 ($1.2\%$)&  1.4928 ($0.9\%$)& 1.4928 ($0.9\%$)\\
     & 6 & 1.6130 & 1.6297 ($1.0\%$)&  1.5929 ($1.2\%$)& 1.5930 ($1.2\%$)\\
     & 7 & 1.7073 & 1.7230 ($0.9\%$)&  1.6788 ($1.7\%$)& 1.6789 ($1.7\%$)\\
     & 8 & 1.7920 & 1.8072 ($0.8\%$)&  1.7537 ($2.1\%$)& 1.7538 ($2.1\%$)\\
    \bottomrule
  \end{tabular}
  \caption{\footnotesize%
    First eight ($n=1,\,2,\,\ldots,\,8$) resonance frequencies
    ($\tilde\omega_n$) for complementary extinction spectrum by distinct
    approaches and computations. Second column: Frequencies
    $\tilde\omega^{\text{eff}}_{n}$ of the homogenization approach.
    Third column: Frequencies $\tilde\omega_{n}^{\text{D}}$ of Dirichlet
    approximation for single nanoribbon. Last two columns: Frequencies
    $\tilde\omega_n$ by direct numerical computation for layered structure
    with $N=4$ and $N=8$ layers. The percentages in parentheses of last
    three columns are the relative deviations of the computed frequencies
    from the respective homogenization results
    $\tilde\omega_{n}^{\text{eff}}$.}
  \label{tab:resonant_frequencies}
\end{table}

%%%%%%%%%%%%%%%%%%%%%%%%%%%%%%%%%%%%%%%%%%%%%%%%%%%%%%%%%%%%%%%%%%%%%%%%%%%%%%%%
%%%%%%%%%%%%%%%%%%%%%%%%%%%%%%%%%%%%%%%%%%%%%%%%%%%%%%%%%%%%%%%%%%%%%%%%%%%%%%%%
%%%%%%%%%%%%%%%%%%%%%%%%%%%%%%%%%%%%%%%%%%%%%%%%%%%%%%%%%%%%%%%%%%%%%%%%%%%%%%%%

\section{Discussion}
\label{sec:discussion}

In this section, we discuss implications of our study in the wave
transmission through plasmonic structures. In particular, we point out the
surprising accuracy of the homogenization approach in regard to the
complementary transmission spectrum; and make an attempt to qualitatively
compare our theoretical predictions to related experimental observations.
Furthermore, we outline a possible extension of our study to the
computation of waveguide modes in layered plasmonic structures with
various sheet geometries.

%%%%%%%%%%%%%%%%%%%%%%%%%%%%%%%%%%%%%%%%%%%%%%%%%%%%%%%%%%%%%%%%%%%%%%%%%%%%%%%%

\subsection{Aspects of homogenized transmission}
\label{subsec:hom-accuracy}

A central question motivating our computations is whether the
homogenization of a layered plasmonic structure with finite thickness can
yield sufficiently accurate results for the wave transmission. In the
homogenization procedure, the error arises from the replacement of
individual layers by a continuous medium through a delicate averaging
process; cf. Eq.~\eqref{eq:eff-general}. The appropriate weight of the
averaging in principle encodes microscale details (e.g., material edges).
This weight is the corrector field.

We have assessed the accuracy of the homogenization results by numerically
computing: (i) the complementary transmission spectrum, $\FTC$, as a
function of frequency; and (ii) a few resonance frequencies,
$\tilde\omega_n$, that characterize the sheet geometry in the case with
nanoribbons. We deem both $\FTC$ and $\tilde\omega_n$ as experimentally
measurable.

In regard to $\FTC$, we observe that the relative homogenization
error remains less than $1\,\%$ for a wide range of frequencies, when the
number, $N$, of layers is as small as $4$. This error may reach a maximum
of about $10\,\%$ near frequencies that correspond to microscale resonances
in the nanoribbon configuration. For fixed $\tilde\omega$, the error
decreases almost inverse linearly with $N$.

The computed resonance frequencies $\tilde\omega_n$
($n=1,\,2,\,\ldots,\,8$) signify the influence on wave transmission of
surface plasmons excited in the nanoribbon geometry. Notably, these
frequencies are accurately captured within the homogenization approach via
the numerical solution of the boundary value problem for the corrector
field. In general, the corresponding relative error ranges from $0.3\,\%$
to $2.1\,\%$ in our numerics. We notice that this error increases with the
order, $n$, of the resonance. This behavior is a manifestation of an
expected limitation of our homogenization approach: The averaging procedure
underlying the homogenized formulas is valid to the leading order in the
spacing $\bar d$, for slowly varying and low-energy incident plane waves.
To achieve the same accuracy for high resonance frequencies ($n\gg 1$), one
would need to properly modify the homogenization procedure. This task is
left for future work. (See also Sec.~\ref{subsec:extensions}).

It is worthwhile to comment on the role of the ENZ condition in the
behavior of the complementary transmission spectrum, $\FTC$, for the
structure with translation invariant sheets; see Fig.~\ref{fig:extinction}.
For fixed wavenumber $k_z$ (tangential to each sheet), the frequency
$\tilde\omega_{\text{ENZ}}$ coming from this condition tends to separate
the frequency axis roughly into two regimes. For $\tilde\omega<
\tilde\omega_{\text{ENZ}}$, the computed $\FTC$ decreases with
$\tilde\omega$ and can have appreciable values less than unity. In this
range, if $\tilde\omega$ is kept fixed while $N$ varies, $\FTC$ is found to
increase with $N$. On the other hand, for $\tilde\omega >
\tilde\omega_{\text{ENZ}}$ (roughly), the computed $\FTC$ may be
non-monotone and tends to become small; thus, wave transmission is
enhanced. In this latter regime ($\tilde\omega >
\tilde\omega_{\text{ENZ}}$), $\FTC$ also loses its monotonicity with
respect to $N$ (for fixed $\tilde\omega$). These observations are
consistent with the expected asymptotic behavior of $\FTC$ for large enough
values of $N$ ($N>100$), not used in our plots. More precisely, in this
limit $\FTC$ should approach a step function with values nearly equal to
unity for $\tilde\omega< \tilde\omega_{\text{ENZ}}$ and close to zero
otherwise. By this large-$N$ picture, the character of the transmitted wave
changes abruptly at the critical value $\tilde\omega =
\tilde\omega_{\text{ENZ}}$: As $\tilde\omega$ decreases, the wave becomes
evanescent below the ``cutoff'' frequency $\tilde\omega_{\text{ENZ}}$
because of the switch in the sign of the emerging effective permittivity.

Interestingly, we find hardly any connection of the ENZ condition to the
extrema of the homogenization error for $\FTC$, in the geometry with
translation invariant sheets (Figs.~\ref{fig:beff} and~\ref{fig:error}).
Because of the absence of microscale (lateral) SPP-related resonances, this
configuration is ideal for examining the frequency dependence of the
homogenization error near the ENZ condition.

In the nanoribbon geometry, the behavior of the \emph{macroscopic} quantity
$\FTC$ versus frequency is a direct consequence of the excitation of
\emph{fine-scale} SPPs on each sheet, because of the presence of material
edges. These SPPs are encoded in the corrector field, $\chi(\vr)$. Near
each resonance frequency, $\chi(\vr)$ has a dominant contribution to the
average of Eq.~\eqref{eq:eff-general}.

%%%%%%%%%%%%%%%%%%%%%%%%%%%%%%%%%%%%%%%%%%%%%%%%%%%%%%%%%%%%%%%%%%%%%%%%%%%%%%%%

\subsection{On the connection of theory to experiment}
\label{subsec:expts}

The setting of our study has been motivated by experiments of wave
transmission through stacks with large, planar graphene sheets and
insulator slabs~\cite{Yanetal2012}. Next, we briefly discuss how trends of
our results can be directly connected to corresponding experimental
observations.

First, the extinction spectrum measured in experiments is a decreasing
function of frequency; see Figs. 2(a), (b) in~\cite{Yanetal2012}. This
behavior is in agreement with the monotonicity of the computed
$\FTC$ for frequencies below $\omega_{\text{ENZ}}$ in the
geometry with translation invariant sheets; see our plots in
Fig.~\ref{fig:extinction}.

Second, for fixed frequency the measured extinction spectrum increases with
the number $N$ of layers in the structure (Figs. 2(a), (b)
in~\cite{Yanetal2012}). This monotonicity with $N$ is also observed in our
computations for $\FTC$ when the sheets are translation
invariant, provided the frequency range is such that $\omega<
\omega_{\text{ENZ}}$ (Fig.~\ref{fig:extinction}).

Furthermore, we make predictions that can possibly be tested in future
experiments with more complicated geometries. In particular, we mention our
prediction of resonances in the wave transmission through the graphene
nanoribbon configuration. An aspect of our computations that deserves
attention for experimental designs is the possible tuning of the (lateral)
SPP resonance through geometric or material parameters of the system. In
this context, we highlight the role of the corrector field. Similar
considerations should hold for other sheet geometries with edges and
microscale defects.

%%%%%%%%%%%%%%%%%%%%%%%%%%%%%%%%%%%%%%%%%%%%%%%%%%%%%%%%%%%%%%%%%%%%%%%%%%%%%%%%
\subsection{Comparison to related homogenization results}
\label{subsec:comparison}

The general homogenization result of \cite{Maier19b}, which provided the main ingredient of our
present study, is primarily concerned with the derivation of an effective material
property in the form of the permittivity tensor $\teff$. This viewpoint is less
concerned with establishing
concrete dispersion relations for propagating modes. In the case with planar
graphene sheets (Sec.~\ref{sec:analysis}), the
homogenization result is compatible with a dispersion relation derived
via Bloch wave theory~\cite{Mattheakisetal2016,Maier18}.

It is instructive to compare our effective permittivity tensor
$\teff$ to the corresponding effective permittivity that underlies the 
\emph{Kronig-Penney model} for plasmonic
crystals~\cite{Halevi1999,Krokhin2010}. In these works, with reference to
the coordinate system of Fig.~\ref{fig:nanoribbons-rve}, an arithmetic average is
used to compute the matrix element $\eff_z$ while a harmonic mean is
applied for $\eff_x$. This result holds true in the long-wavelength limit
which is equivalent to our assumption that $\beta d\to 0$. The
homogenization result is compatible with our Eq.~\eqref{eqs:hom-eta} if the
sheets are translation invariant. In this case the harmonic and arithmetic
means give the same result provided that the permittivity and conductivity
are spatially constant. More generally, however, the main difference between
the two results is that the role of the harmonic mean is replaced by a
weighted arithmetic average governed by the corrector field,
$\chi_{\para}$, in our formalism. For example, the picture drastically
changes from the situation of translation invariant sheets when one
introduces fine-scale lateral SPP resonances that do not vanish in the
long-wavelength limit. This is the case for the nanoribbon configuration
(Secs.~\ref{sec:homogen-reson} and \ref{sec:computation}): the resonances
of Fig.~\ref{fig:ribbons-extinction} are solely caused by the corrector
field and cannot be captured directly by a harmonic average.

%%%%%%%%%%%%%%%%%%%%%%%%%%%%%%%%%%%%%%%%%%%%%%%%%%%%%%%%%%%%%%%%%%%%%%%%%%%%%%%%
\subsection{Stability of the homogenization result under random perturbations}
\label{subsec:stability}

An important question concerns the stability of the homogenized Fresnel
coefficients under some random variation of the problem parameters. A
complete or rigorous answer to this question lies beyond our present scope.
We are tempted, however, to present some brief heuristic arguments that
address some notable cases. We mostly restrict our discussion to the
stability of the entries of the effective permittivity tensor $\teff$. The
stability of related observable quantities such as the Fresnel coefficients
can in principle be assessed from their respective formulas by applying the
normal distribution for the error of $\teff$. In this vein, we do not
expect an error amplification (cf. Eqs.~\eqref{eq:homogenized_general}
and~\eqref{eq:beff}). We also do not address any homogenization errors
beyond the finite-size effects that were already studied numerically in
Sec.~\ref{sec:computation}.

Recall that the effective permittivity tensor $\teff$ is given by
Eq.~\eqref{eq:eff-general}. We heuristically distinguish two types
of perturbations: (i) Random variations that mainly influence the averaging
procedure and change the corrector contribution negligibly; and (ii) random
variations that change the corrector contribution appreciably. In the
present setting of nanoscale resonators (conducting sheets intercalated in
dielectrics), examples for type (i) include random variations of
periodicity, alignment and orientation that keep the shape and dimensions
of the nanoscale resonator intact. In this case, we expect that the
homogenization procedure is well controlled, and thus stable.

The situation is different for random variations of type (ii), which modify
essential geometric features of the resonators. Examples for this type
include random variations of the ribbon width, or the Drude weight of the
surface conductivity. By revisiting the Dirichlet
approximation~\eqref{eq:dirichlet_approximation} and
Table~\ref{tab:resonant_frequencies}, we expect that the homogenization
procedure remains stable as long as random variations for length scales
under type (ii) are sufficiently smaller (in some sense) than the SPP
wavelength: Then resonance frequencies will only shift slightly. Hence, it
is reasonable to expect that random variations will only lead to slight
broadening of the computed, unperturbed resonance behavior. (Regarding the
complementary transmission spectrum, the unperturbed behavior is shown in
Figs.~\ref{fig:ribbons-extinction} and \ref{fig:ribbons-error}).

%%%%%%%%%%%%%%%%%%%%%%%%%%%%%%%%%%%%%%%%%%%%%%%%%%%%%%%%%%%%%%%%%%%%%%%%%%%%%%%%

\subsection{Extension: Waveguide modes in plasmonic heterostructures}
\label{subsec:extensions}

Our computational framework can be used to determine waveguide modes
supported by plasmonic crystals of flat sheets at any given frequency,
$\omega$. The objective is to obtain the wavenumber $k_z$ as a function of
$\omega$ (see Figs.~\ref{fig:sandwich} and~\ref{fig:nanoribbons-rve}). In
principle, a multitude of such modes may exist for some fixed $\omega$.

This problem is equivalent to searching for certain types of singularities
(poles) in the Fresnel coefficients $R$ and $T$ as functions of the, in
principle complex, variable $k_z$. A related question is how the resulting
wavenumbers $k_z$ of the layered structure compare to their counterparts of
the homogenized crystal. An appeal of our formalism is the incorporation of
geometries other than translation invariant sheets via the corrector field;
cf.~\cite{Mahmoodi2019}.

To illustrate some of the technical aspects of determining $k_z(\omega)$,
we turn our attention to Eqs.~\eqref{eq:fresnel}
and~\eqref{eq:homogenized_general} which pertain to the Fresnel
coefficients of the layered and homogenized structure, respectively. By
setting equal to zero the denominator of Eq.~\eqref{eq:fresnel}, we obtain
the dispersion relation
\begin{align*}
  \eb_{\text{s}+}\,t_{1+}-\eb_{\text{s}-}\,t_{2+}\,e^{\im2\beta d}=0~.
\end{align*}
The admissible solutions, $k_z^{\text{lr}}(\omega)$, of this equation yield
the lateral waveguide modes of the full layered structure. On the other
hand, the homogenized dispersion relation is
\begin{align*}
  \eb_{\text{s}+}\,\eb_{0+}-\eb_{\text{s}-}\,\eb_{0-}\,e^{\im 2\beff H}=0,
\end{align*}
with possible solutions $k_z^{\text{hm}}(\omega)$. The question is how the
values for $k_z^{\text{lr}}(\omega)$ compare to those for
$k_z^{\text{hm}}(\omega)$. This question can be addressed numerically for
distinct geometries. A detailed study lies beyond our present scope.

It turns out that the answer to the above question depends on the order of
the mode, i.e., how high the value of $|k_z|$ needs to be for fixed
$\omega$. Because our homogenization procedure is valid for slowly varying
and low-energy waves, it can provide reasonably accurate results for low
values of $|k_z|$. In contrast, our homogenized results may become
questionable for sufficiently large $|k_z|$. The situation regarding the
homogenization error for $k_z$ (for fixed $\omega$) versus the mode number
is analogous to that for $\tilde\omega_n$ versus $n$ depicted in the last
two columns of Table~\ref{tab:resonant_frequencies}.

These considerations point to the need for extending the homogenization
procedure to high values of $|k_z|$. In the transfer matrix approach, for
translation invariant sheets, we can seek this extension by relaxing the
main assumptions of Sec.~\ref{sec:small_spacing}. More precisely, in the
limit as $k_0 d\to 0$ we should assume that
\begin{align*}
  |\beta d|\simeq {\rm const.},\quad
  \left|\frac{\omega\mu\sigma}{\beta}\right|\ll 1,
  \quad
  \sigma\diagup d\simeq {\rm const.}.
\end{align*}
The first condition now replaces the previously applied statement $|\beta
d|\ll 1$. The resulting expansions for $R$ and $T$ are deemed as manageable
in this case.

For more general geometries, when the transfer matrix approach is not
directly applicable as above, the homogenization procedure needs a delicate
modification to accommodate high values of $|k_z|$. In this case, microscale
details of the conducting sheets are incorporated into the appropriate
(nontrivial) corrector field. This problem is the subject of work in
progress.

%%%%%%%%%%%%%%%%%%%%%%%%%%%%%%%%%%%%%%%%%%%%%%%%%%%%%%%%%%%%%%%%%%%%%%%%%%%%%%%%
%%%%%%%%%%%%%%%%%%%%%%%%%%%%%%%%%%%%%%%%%%%%%%%%%%%%%%%%%%%%%%%%%%%%%%%%%%%%%%%m
%%%%%%%%%%%%%%%%%%%%%%%%%%%%%%%%%%%%%%%%%%%%%%%%%%%%%%%%%%%%%%%%%%%%%%%%%%%%%%%m

\section{Conclusion}
\label{sec:conclusion}

In this paper, we studied analytically and numerically the wave
transmission through plasmonic crystals for a wide range of frequencies.
These structures consist of a \emph{finite number} of mutually parallel
conducting sheets intercalated between dielectric hosts, and are
practically appealing. We computed the associated Fresnel coefficients by
two alternate approaches. One method relies on direct computations for the
full layered structure. Another approach makes use of homogenization, i.e.,
the replacement of the individual layers by an appropriately determined, in
principle anisotropic, continuous medium in a slab of equal total
thickness. We considered the two distinct geometries with infinite,
translation invariant sheets and nanoribbons.

Our results indicate the very good accuracy of the homogenized formula for
the complementary transmission spectrum even for a relatively small number
of layers. This result is a highlight of our analysis and numerics. In the
case with nanoribbons, the maximum relative error occurs at SPP-related
resonance frequencies. Notably, the first few (lowest) frequencies are
captured accurately by the \emph{corrector field} of the homogenization
procedure which enters the effective permittivity tensor. This field
incorporates effects from details of the microscale geometry, e.g., edges
and defects. The use of the corrector in accurately identifying lateral
resonances in layered plasmonic structures with finite total thickness is
another noteworthy ingredient of our work.

A couple of open problems inspired by our work should be mentioned. It
would be of interest to extend the computations to other configurations of
possible experimental relevance such as those with flat microdisks or
curved sheets. Another open task is that of a thorough stability analysis
of the homogenization results under random variations of the problem
parameters that would quantify the heuristic arguments of
Sec.~\ref{subsec:stability}. The homogenization procedure described here is
applicable to low enough wavenumbers. It must be properly modified to
accurately capture effects of high wavenumbers. This issue arises in the
study of wave\-guide modes allowed to propagate through the plasmonic
crystal. This problem deserves some attention.

%%%%%%%%%%%%%%%%%%%%%%%%%%%%%%%%%%%%%%%%%%%%%%%%%%%%%%%%%%%%%%%%%%%%%%%%%%%%%%%m
%%%%%%%%%%%%%%%%%%%%%%%%%%%%%%%%%%%%%%%%%%%%%%%%%%%%%%%%%%%%%%%%%%%%%%%%%%%%%%%%
%%%%%%%%%%%%%%%%%%%%%%%%%%%%%%%%%%%%%%%%%%%%%%%%%%%%%%%%%%%%%%%%%%%%%%%%%%%%%%%%

\begin{acknowledgements}
  We wish to thank Professors M.~I. Weinstein, A. Al\`{u} and P. Cazeaux
  for engaging discussions on homogenization and metamaterials. One of us
  (D.M) also thanks Professor T.~T. Wu for a discussion which motivated
  part of this work. The authors acknowledge partial support by the ARO
  MURI award W911NF-14-1-0247. M.M. also acknowledges partial support by
  the NSF under grant DMS-1912847. M.L. was also supported by the NSF under
  grant DMS-1906129. The research of D.M. was also partially supported by a
  Research and Scholarship award by the Graduate School, University of
  Maryland, in the spring of 2019. Part of this research was carried out
  when the authors were visiting the Institute for Pure and Applied
  Mathematics (IPAM), which is supported by NSF under grant DMS-1440415.
\end{acknowledgements}

%%%%%%%%%%%%%%%%%%%%%%%%%%%%%%%%%%%%%%%%%%%%%%%%%%%%%%%%%%%%%%%%%%%%%%%%%%%%%%%%
%%%%%%%%%%%%%%%%%%%%%%%%%%%%%%%%%%%%%%%%%%%%%%%%%%%%%%%%%%%%%%%%%%%%%%%%%%%%%%%%
%%%%%%%%%%%%%%%%%%%%%%%%%%%%%%%%%%%%%%%%%%%%%%%%%%%%%%%%%%%%%%%%%%%%%%%%%%%%%%%%

%%%%%%%%%%%%%%%%%%%%%%%%%%%%%%%%%%%%%%%%%%%%%%%%%%%%%%%%%%%%%%%%%%%%%%%%%%%%%%%%
%%%%%%%%%%%%%%%%%%%%%%%%%%%%%%%%%%%%%%%%%%%%%%%%%%%%%%%%%%%%%%%%%%%%%%%%%%%%%%%%
%%%%%%%%%%%%%%%%%%%%%%%%%%%%%%%%%%%%%%%%%%%%%%%%%%%%%%%%%%%%%%%%%%%%%%%%%%%%%%%%

\appendix
\section{Details of transfer matrix approach}
\label{app:transfer_matrix}

In this appendix, we provide details for solving the full transmission problem
of Sec.~\ref{sec:analysis} via the transfer matrix method. The incoming plane
wave has TM polarization. First, we review the derivation of the dispersion
relation $\beta(k_z)$ for the wavenumber in the $x$-direction in the
anisotropic medium of dielectric permittivity $\te(\omega)$. Second, we sketch
the steps for deriving explicit formulas for the Fresnel coefficients $R(k_z)$
and $T(k_z)$ that characterize the plane wave propagation through the layered
structure of Fig.~\ref{fig:sandwich}. Our approach is based on applying a
cascade of elementary transmission problems.

%%%%%%%%%%%%%%%%%%%%%%%%%%%%%%%%%%%%%%%%%%%%%%%%%%%%%%%%%%%%%%%%%%%%%%%%%%%%%%%%

\subsection{Derivation of dispersion relation $\beta(k_z)$}
\label{appsub:dispersion}

Let us first describe the dispersion of a TM-polarized plane wave propagating
in a homogeneous anisotropic medium of permittivity
$\te=\text{diag}\,(\e_x,\e_y,\e_z)$. Consider the following ansatz for the
electric field:
\begin{align*}
  \vE(\vr)= {\boldsymbol \ME}\,e^{i\vk\cdot\vr},
\end{align*}
where $\vk=(k_x,k_y,k_z)$, $\vr=(x,y,z)$ and $\boldsymbol \ME$ is a constant
vector. Maxwell's equations imply the statement
\begin{align*}
  \vk\times(\vk\times\boldsymbol\ME) +
\omega^2\mu\underline{\e}\,\boldsymbol\ME\;=\;0,
\end{align*}
which must be solved for nonzero $\boldsymbol\ME$.

For TM polarization, we write $\boldsymbol\ME=(\ME_x,0, \ME_z)$ and
$\vk=(\beta,0, k_z)$. Hence, the above equation yields the system
\begin{align*}
  \begin{pmatrix}
    \omega^2\mu\e_\perp\,-\,k_z^2 & \beta k_z
    \\
    \beta k_z & \omega^2\mu\e_\para\,-\,\beta^2
  \end{pmatrix}
  \begin{pmatrix}
    \ME_x
    \\
    \ME_z
  \end{pmatrix}
  \;=\;0
\end{align*}
where $\e_\perp=\e_x$ and $\e_\para=\e_z$.
This linear system has a \emph{nontrivial} solution $(\ME_x,\ME_z)$ if
\begin{align*}
  \big(\omega^2\mu\e_\perp\,-\,k_z^2\big)
  \big(\omega^2\mu\e_\para\,-\,\beta^2\big)
  \;-\;
  \beta^2k_z^2
  \;=\;0
\end{align*}
which yields Eq.~\eqref{eq:kx-beta-disp} for $\beta(k_z)$ (see
Sec.~\ref{sec:analysis}).

%%%%%%%%%%%%%%%%%%%%%%%%%%%%%%%%%%%%%%%%%%%%%%%%%%%%%%%%%%%%%%%%%%%%%%%%%%%%%%%%

\subsection{Two elementary transmission problems}
\label{appsub:elem-probs}

Next, we study two basic transmission problems in order to simplify the
analysis for the full layered structure by the transfer matrix approach. These
problems are: (i) Propagation into an anisotropic dielectric of permittivity
$\te$ by a given distance; and (ii) transmission through a sheet of surface
conductivity $\sigma$.

\paragraph*{(i) Propagation in dielectric host by distance $L$.}
For algebraic convenience, we use the $x$-dependent part, $\ME(x)$, of the
$z$-component, $E_z(x,z)$, of the electric field from
Sec.~\ref{sec:analysis}. At position $x=x_1$, this $\ME(x)$ has the form
\begin{align*}
  \mathcal{E}(x=x_1)=A^{-}e^{-\im\beta x_1}+C^{-}e^{\im\beta x_1}.
\end{align*}
Hence, at position $x=x_2=x_1+L$ we have
\begin{equation*}
  \mathcal{E}(x=x_2) = A^{+}e^{-\im\beta x_1}+C^{+}e^{\im\beta x_1}
\end{equation*}
where
\begin{align}
  \label{eq:TI-def}
  \begin{pmatrix}
    A^{+} \\ C^{+}
  \end{pmatrix}
  =
  \;\mathcal{T}_{\text{I}}(L)\;
  \begin{pmatrix}
    A^{-} \\ C^{-}
  \end{pmatrix},
  \quad
  \mathcal{T}_{\text{I}}(L)\;=\;
  \begin{pmatrix}
    e^{-\im\beta L} & 0 \\ 0 & e^{\im\beta L}
  \end{pmatrix}.
\end{align}
Recall that $\beta$ is given by Eq.~\eqref{eq:kx-beta-disp}.

\paragraph*{(ii) Transmission through conducting sheet.}
Suppose that a sheet with conductivity $\sigma$ is situated at $x=0$,
between two dielectrics of permittivity $\te$. Let us take
\begin{equation*}
  \mathcal{E}(x)=A^{-}e^{-\im\beta x} +
  C^{-}e^{\im\beta x},\quad x<0.
\end{equation*}
After transmission through the sheet, the related amplitudes change; thus, we have
\begin{equation*}
  \mathcal{E}(x) = A^{+}c^{-\im\beta x} + C^{+}e^{\im\beta x},\quad x>0.
\end{equation*}
We need to describe the matrix, $\mathcal T_{\text{II}}$, that connects
$(A^+, C^+)$ and $(A^-,C^-)$.

For this purpose, let us consider the boundary conditions obeyed by the
electromagnetic field across the sheet. First, $E_z(x,z)$ must be
continuous at $x=0$~\cite{Maier17}. Thus, we impose the condition
\begin{equation}\label{eq:E-cont}
 \mathcal{E}(0^+)- \mathcal{E}(0^-)=0,
 \end{equation}
where the statement $x=0^+$ ($x=0^-$) means that $x$ approaches $0$ from
above (below). In addition, the surface current density induced on the
sheet causes a jump on the tangential component, $B_y$, of the magnetic
field at $x=0$. By use of Maxwell's equations we can express $B_y$ in terms
of $E_z$. Thus, we require that
\begin{equation}\label{eq:B-jump}
  \frac{\im\omega\e_\perp}{k_\perp^2-k_z^2}
  \,\left\{\frac{\text{d}\mathcal{E}(0^+)}{\text{d}x}
  \,-\,\frac{\text{d}\mathcal{E}(0^-)}{\text{d}x}\right\}\;=\;\sigma
  \mathcal{E}(0)
\end{equation}
where $k_\perp^2=\omega^2\mu\e_\perp$ (see Sec.~\ref{sec:analysis}).

By replacing $\ME(x)$ in Eqs.~\eqref{eq:E-cont} and~\eqref{eq:B-jump} by
the formulas involving $A^{\pm}$ and $C^{\pm}$, we obtain the relation
\begin{equation*}
  \begin{pmatrix} A^{+} \\ C^{+} \end{pmatrix}
  =\mathcal{T}_{\text{II}} \begin{pmatrix} A^{-} \\ C^{-}\end{pmatrix}
\end{equation*}
where
\begin{align}\label{eq:TII-def}
  \qquad
  \mathcal{T}_{\text{II}}\;=\;
  \begin{pmatrix}
    1+\frac{\omega\mu\sigma}{2\beta}\,\frac{k_\perp^2-k_z^2}{k_\perp^2}
    &
    \frac{\omega\mu\sigma}{2\beta}\,\frac{k_\perp^2-k_z^2}{k_\perp^2}
    \\
    -\frac{\omega\mu\sigma}{2\beta}\,\frac{k_\perp^2-k_z^2}{k_\perp^2}
    &
    1-\frac{\omega\mu\sigma}{2\beta}\,\frac{k_\perp^2-k_z^2}{k_\perp^2}
  \end{pmatrix}.
\end{align}

For later algebraic convenience, we choose to rewrite the above expression
for transfer matrix $\mathcal{T}_{\text{II}}$ in an alternate form by using
the {\em effective} wavenumber
\begin{align*}
  \beff = \beta\sqrt{1 +\im\, \frac{\omega\mu\sigma}{\beta}\, \frac{1}{\beta d} \,
  \frac{k^2_\perp - k_z^2} {k^2_\perp}}.
\end{align*}
Recall that $d$ denotes the interlayer spacing (see
Fig.~\ref{fig:sandwich}). The above definition of $\beff$ is motivated by
the homogenization procedure of Sec.~\ref{sec:homogen-reson}, where the
ratio $\sigma/d$ is treated as a ($d$-independent) constant if $|\beta
d|\ll 1$; see~\cite{Maier19b,Maier18}. Accordingly, we obtain the
expression
\begin{align*}
  \mathcal{T}_{\text{II}} &=
  \begin{bmatrix}
    1-\frac{id}{2\beta}\big((\beta^{\text{eff}})^2 - \beta^2\big)
    & -\frac{id}{2\beta}\big((\beta^{\text{eff}})^2 - \beta^2\big)
    \\
    \frac{id}{2\beta}\big((\beta^{\text{eff}})^2 - \beta^2\big)
    & 1+\frac{id}{2\beta}\big((\beta^{\text{eff}})^2 - \beta^2\big)
  \end{bmatrix}.
\end{align*}

%%%%%%%%%%%%%%%%%%%%%%%%%%%%%%%%%%%%%%%%%%%%%%%%%%%%%%%%%%%%%%%%%%%%%%%%%%%%%%%%

\subsection{Transmission through full multilayer system}
\label{appsub:full}

Next, we sketch a derivation for the Fresnel coefficients $R(k_z)$ and
$T(k_z)$ of the full layered structure (see Fig.~\ref{fig:sandwich}). Our
procedure relies on the successive application of results for the
elementary problems (i) and (ii) stated above.

Let the field $\ME(x)$ in a slab of the layered structure, for
$(n-1)d<x<nd$, be described by
\begin{equation*}
  \ME(x)=A^{(n)}\, e^{-\im k_x x}+C^{(n)}\,e^{\im k_x x},\, k_x=\beta;\,n=1,\,\ldots,\,N.
\end{equation*}
From Sec.~\ref{appsub:elem-probs}, the amplitudes $A^{(n)}$ and $C^{(n)}$ satisfy
\begin{equation*}
  \begin{pmatrix}
    A^{(n+1)} \\ C^{(n+1)}
  \end{pmatrix}
  =\{\mathcal{T}_{\text{I}}(L)\}^{-1}\mathcal T_{\text{II}} \mathcal
  T_{\text{I}}(L)
  \begin{pmatrix}
    A^{(n)} \\ C^{(n)}
  \end{pmatrix},\quad L=nd,
\end{equation*}
if $n=1,\,\ldots,\,N-1$. Note the identity $\mathcal
T_{\text{I}}(md)=\{\mathcal T_{\text{I}}(d)\}^m$ for any integer $m$. By
successively applying the above recursive relation for $(A^{(n)},
C^{(n)})$, we obtain
\begin{align}
  \begin{pmatrix}
    A^{(N)} \\ C^{(N)}
  \end{pmatrix}
  =\mathcal{T}
  \begin{pmatrix}
    A^{(1)} \\ C^{(1)}
  \end{pmatrix},
  \;
  \mathcal{T} = \mathcal{T}_{\text{I}}^{1-N}\,\left(\mathcal{T}_{\text{II}}
  \mathcal{T}_{\text{I}}\right)^{N-1}
  \label{eq:transm}
\end{align}
where the symbol $\mathcal T_{\text{I}}$ stands for $\mathcal T_{\text{I}}(d)$.

A remark on the analytical computation of the matrix $\mathcal{T}$ is in
order. This calculation is carried out via the diagonalization of
$\mathcal{T}_{\text{II}} \mathcal{T}_{\text{I}}$ (see also
Appendix~\ref{app:transfer_matrix_expansion}). Accordingly, we explicitly
determine the invertible matrix $\mathcal S$ such that
$\mathcal{T}_{\text{II}} \mathcal{T}_{\text{I}}=\mathcal
S\,\text{diag}(\lambda_+, \lambda_-)\,\mathcal S^{-1}$ where $\lambda_\pm$
are the eigenvalues of $\mathcal{T}_{\text{II}} \mathcal{T}_{\text{I}}$.
Clearly, this $\mathcal S$ is formed by eigenvectors of
$\mathcal{T}_{\text{II}} \mathcal{T}_{\text{I}}$, which are calculable in
closed form. Hence, we write
\begin{equation*}
  \mathcal{T}=\mathcal T_{\text{I}}^{1-N}\mathcal S\,
  \text{diag}\big(\lambda_+^{N-1}, \lambda_-^{N-1}\big)\,
  \mathcal{S}^{-1}
\end{equation*}
where $\mathcal T_{\text{I}}^{1-N}=\mathcal T_I((1-N)d)$. The formulas for
$\mathcal S$ and $\lambda_{\pm}$ as well as an ensuing approximation for
$\mathcal{T}$ if $N$ is large and $\beta d\to 0$ are discussed in
Appendix~\ref{app:transfer_matrix_expansion}.

The remaining task here is to relate the amplitudes $A^{(n)}$ and $C^{(n)}$
for $n=1$ and $n=N$ to coefficients $R$ and $T$. Therefore, we need to
apply the suitable transmission conditions at the corresponding interface
between dielectric host and air or substrate, at $x=0$ or $x=Nd=H$.

First, we consider the interface at $x=0$. Recall that
\begin{align*}
  \ME(x)=\ME^\air(x) = R\,e^{-\im k_{x,0}\,x}+e^{\im k_{x,0}\,x},\,x<0,
\end{align*}
while $\ME(x)=A^{(1)} e^{-\im \beta x}+C^{(1)} e^{\im\beta x}$ if $0<x<d$.
By requiring that the field components $E_z(x,z)$ and $B_y(x,z)$ be
continuous at $x=0$, we find that
\begin{align*}
  \begin{cases}
    \begin{aligned}
      A^{(1)} + C^{(1)} &= 1 + R,
      \\
      \frac{\im\omega\mu\e_\perp}{k_\perp^2-k_z^2}
      (\im\beta)\big(-A^{(1)}+C^{(1)}\big)&=
      \frac{\im\omega\mu\e_{0}}{k_{x,0}^2}
      (\im k_{x,0})
      \big(-R+1\big).
    \end{aligned}
  \end{cases}
\end{align*}
This system yields
\begin{align}\label{eq:A-C-R}
  \begin{cases}
    \begin{aligned}
      A^{(1)}&=\frac{1}{2}\Big[1+R-\frac{\e_0}{\e_\perp}\frac{k_\perp^2-k_z^2}{k_{x,0}\beta}(1-R)\Big],\\
      C^{(1)}&=\frac{1}{2}\Big[1+R+\frac{\e_0}{\e_\perp}\frac{k_\perp^2-k_z^2}{k_{x,0}\beta}(1-R)\Big].
    \end{aligned}
  \end{cases}
\end{align}

Similarly, consider the interface at $x=H$. The transmitted wave is
\begin{align*}
  \ME(x)=\ME^\sub(x) = T\,e^{\im k_{x,\text{s}}\,x},\ x>H=Nd,
\end{align*}
while $\ME(x)=A^{(N)} e^{-\im\beta x}+C^{(N)} e^{\im\beta x}$ for
$(N-1)d<x<H$. Accordingly, we obtain the system
\begin{align*}
  \begin{cases}
    \begin{aligned}
      A^{(N)}\,e^{-\im\beta Nd} + C^{(N)}\,e^{\im\beta ND} \;=\;
      T\,e^{i k_{x,\text{s}}Nd},
      \\[0.5em]
      \frac{\im\omega\mu\e_\perp}{k_\perp^2-k_z^2}
      (\im\beta)\big(-A^{(N)}\,e^{-\im\beta Nd}+C^{(N)}\,e^{\im\beta ND}\big)
      \qquad\qquad
      \\
      =
      \frac{\im\omega\mu\e_{\text{s}}}{k_{x,\text{s}}^2}
      (\im k_{x,\text{s}})
      T\,e^{i k_{x,\text{s}}Nd}
    \end{aligned}
  \end{cases}
\end{align*}
which entails
\begin{align}\label{eq:A-C-T}
  \begin{cases}
    \begin{aligned}
      A^{(N)}&=\frac{T}{2}e^{\im (\beta+k_{x,\text{s}})H}\Big(1-\frac{\e_{\text{s}}}{\e_\perp}\frac{k_\perp^2-k_z^2}{\beta k_{x,\text{s}}}\Big),\\
      C^{(N)}&=\frac{T}{2}e^{\im(-\beta+k_{x,\text{s}})H}\Big(1+\frac{\e_{\text{s}}}{\e_\perp}\frac{k_\perp^2-k_z^2}{\beta k_{x,\text{s}}}\Big).
    \end{aligned}
  \end{cases}
\end{align}
After some algebra, Eqs.~\eqref{eq:transm}--\eqref{eq:A-C-T} yield the
Fresnel coefficients $R(k_z)$ and $T(k_z)$. The resulting formulas are
displayed compactly in Eq.~\eqref{eq:fresnel} (Sec.~\ref{sec:analysis}).

%%%%%%%%%%%%%%%%%%%%%%%%%%%%%%%%%%%%%%%%%%%%%%%%%%%%%%%%%%%%%%%%%%%%%%%%%%%%%%%%
%%%%%%%%%%%%%%%%%%%%%%%%%%%%%%%%%%%%%%%%%%%%%%%%%%%%%%%%%%%%%%%%%%%%%%%%%%%%%%%%
%%%%%%%%%%%%%%%%%%%%%%%%%%%%%%%%%%%%%%%%%%%%%%%%%%%%%%%%%%%%%%%%%%%%%%%%%%%%%%%%

\section{Derivation of homogenized system via transfer matrix approach}
\label{app:transfer_matrix_expansion}

In this appendix, we outline the derivation of the homogenized Fresnel
coefficients $R^{\text{eff}}$ and $T^{\text{eff}}$ in the limit of small
interlayer spacing $d$, as $\beta d\to 0$. To this end, we use the
framework of Appendix~\ref{app:transfer_matrix}.

Accordingly, we can write
\begin{align*}
  \mathcal{T}_{\text{II}} \mathcal{T}_{\text{I}}
  \;=\; \mathcal S
  \begin{pmatrix}
    \lambda_+ & 0 \\ 0 & \lambda_-
  \end{pmatrix}
  \mathcal S^{-1},
\end{align*}
where $\mathcal S$ is a suitable non-singular matrix and $\lambda_\pm$ are
the two eigenvalues of $\mathcal{T}_{\text{II}} \mathcal{T}_{\text{I}}$. A
direct computation for $\lambda_\pm$ under the conditions
\begin{align*}
  |\beta d|\ll 1,\quad
  \left|\frac{\omega\mu\sigma}{\beta}\right|\ll 1,
  \quad
  \sigma\diagup d\simeq {\rm const.},
\end{align*}
yields the formula
\begin{align*}
  \lambda_\pm \;\simeq\;
  e^{\pm\im\beff d};
  \quad
  \beff = \sqrt{\beta^2 + \frac{\im\omega\mu\sigma} {d} \,
  \frac{k^2_\perp - k^2_z} {k^2_\perp}}.
\end{align*}
Recall that the effective wavenumber $\beff$ was introduced in
Eq.~\eqref{eq:beta-eff-def} in an ad hoc fashion. After some algebra, the
transformation matrix $\mathcal S$ can be written as
\begin{align*}
  \mathcal S=
  \begin{pmatrix}
    \check\beta\,e^{\im\beta d} & \check\beta\,e^{\im\beta d} \\
    \lambda_+ - (1+\check\beta)e^{\im\beta d} & \lambda_- - (1+\check\beta)e^{\im\beta
    d}
  \end{pmatrix},
\end{align*}
where $\check\beta=\frac{\im d}{2\beta}\big\{\beta^2-(\beff)^2\big\}$. A
small-$d$ expansion for
\begin{align*}
  \begin{bmatrix}
    t_{11} & t_{12} \\ t_{21} & t_{22}
  \end{bmatrix}
  \;=\;
  \big[\mathcal{T}_{\text{II}} \mathcal{T}_{\text{I}}\big]^{N-1}
  \;=\; \mathcal S
  \begin{pmatrix}
    \lambda_+^{N-1} & 0 \\ 0 & \lambda_-^{N-1}
  \end{pmatrix}
  \mathcal S^{-1}
\end{align*}
furnishes the formulas
\begin{align*}
  t_{11}\;&\simeq\; \cos(\beff H)\;-\;\im\,\frac{\beta^2+(\beff)^2}{2\beta\beff}\,\sin(\beff H),
  \\[0.1em]
  t_{12}\;=\;-t_{21}\;&\simeq\;
  -\,\im\,\frac{(\beff)^2-\beta^2}{2\beta\beff}\,\sin(\beff H),
  \\[0.1em]
  t_{22}\;&\approx\; \cos(\beff H)\;+\;\im\,\frac{\beta^2+(\beff)^2}{2\beta\beff}\,\sin(\beff H).
\end{align*}
The homogenized coefficients of Eq.~\eqref{eq:homogenized_layers} are
obtained by substitution of the above approximations for $t_{ij}$ into
Eq.~\eqref{eq:fresnel}; and the simplifications  $k_z=0$ and $\e_{\text{s}}
= \e_0$.

%%%%%%%%%%%%%%%%%%%%%%%%%%%%%%%%%%%%%%%%%%%%%%%%%%%%%%%%%%%%%%%%%%%%%%%%%%%%%%%%
%%%%%%%%%%%%%%%%%%%%%%%%%%%%%%%%%%%%%%%%%%%%%%%%%%%%%%%%%%%%%%%%%%%%%%%%%%%%%%%%

\section{General periodic homogenization}
\label{app:hom-gen-th}

In this appendix, we review the general homogenization result for periodic 
plasmonic crystals~\cite{Maier19b}. The underlying methodology relies on two-scale asymptotic
expansions for solutions of the time-harmonic Maxwell equations. The main
assumptions can be stated as follows. Firstly, the material has a suitable periodic
microstructure. This means that material parameters such as the
permittivity of the dielectric host and the surface conductivity of the
sheets have a well-defined microscopic periodicity which can be expressed
via a representative volume element (cell); see
Fig.~\ref{fig:nanoribbons-rve}. This element is repeated periodically in
all spatial directions; in particular, in the $x$-direction its length is
$d$ which signifies the microscale size.

Secondly, a \emph{separation of length scales} has to occur. In this
regard, recall the hypotheses of Sec.~\ref{sec:small_spacing}. The main
assumption of scale separation is that the wavelength of plane wave
propagation in air (with wavenumber $k_0$) is much larger than the length
scale of the TM-polarized SPP on the single sheet. The latter length scales
linearly with $\sigma$~\cite{Lowetal2017,Maier17} and should be of the
order of $d$ here. Hence, we apply the familiar conditions (see
Sec.~\ref{sec:small_spacing})
\begin{align*}
  |\beta d|\ll 1,
  \quad
  \left|\frac{\omega\mu\sigma}{\beta}\right|\ll 1,
  \quad
  \sigma\diagup d\simeq {\rm const}.
\end{align*}
Consequently, the layered plasmonic crystal can be replaced by an
appropriate continuous anisotropic medium which has an effective
permittivity tensor, $\teff$~\cite{Maier19b}.  The formula for this $\teff$
in principle contains: (i) a weighted bulk average of the permittivity
$\te$ of the dielectric host; and (ii) a similarly weighted surface average
of the conductivity $\sigma$ of the 2D material. The weights for these
averages depend on a (local) fine-scale, vector valued corrector field,
$\boldsymbol\chi$, defined in the representative volume
element (Fig.~\ref{fig:nanoribbons-rve})~\cite{Maier19b}. This $\boldsymbol\chi$ satisfies a boundary value
problem in the representative volume element; the boundary conditions
account for wave transmission through the arbitrarily shaped conducting
sheet. In the present case, where $\te=\text{diag}(\e,\e,\e)$ describes the permittivity microstructure, the bulk
average contribution trivially reduces to $\te$~\cite{Maier19b}. In
contrast, the surface average can be complicated for arbitrary sheet
geometry. More precisely, we obtain the following formula for the
effective-permittivity matrix elements~\cite{Maier19b}:
\begin{align}\label{eq:eff-tensor-app}
  \eff_{ij} \;=\; \varepsilon\delta_{ij} -\frac{\sigma(\omega)}{\im\omega
  d^3}\int_{\Sigma}
  \left[\boldsymbol\tau_j+\nabla_{\boldsymbol\tau}\chi_j(\vr)\right]
  \cdot\boldsymbol e_i\,\db\vr
\end{align}
where $i,\,j=x,\,y,\,z$; cf. Eq.~\eqref{eq:eff-general}.
Each scalar field $\chi_i(\vr)$ is a potential-type
function that is periodic in the representative volume element and encodes
features (e.g., edges) of the sheet geometry. Therefore, $\chi_i$ captures
fine-scale lateral plasmonic resonances that are possibly excited in the 2D
material. By asymptotics, we have obtained the governing differential
equation and associated boundary conditions for
$\chi_i(\vr)$~\cite{Maier19b}. In particular, in the interior of the
representative volume element but outside the sheet $\Sigma$, the field
$\chi_i(\vr)$ solves the Laplace equation, viz.,
\begin{equation}\label{eq:chi-pde}
  \Delta\chi_i(\vr)=0;
\end{equation}
$\Delta=\partial^2/\partial x^2+ \partial^2/\partial y^2+
\partial^2/\partial z^2 $ denotes the Laplacian.

In addition, $\chi_i$ obeys two transmission conditions across $\Sigma$.
First, the field $\chi_i(\vr)$ must be continuous across $\Sigma$; thus,
its tangential derivatives also are. This continuity condition ensures that
the tangential electric field is continuous on $\Sigma$. Second, the normal
derivative of $\chi_i$ has a jump discontinuity proportional to $\sigma$
across $\Sigma$, viz.,
\begin{align}\label{eq:chi-bc}
  \boldsymbol\nu\cdot [(\nabla\chi_i)^{+}-(\nabla\chi_i)^{-}]=
  \frac{\sigma(\omega)}{\im\omega\e} \nabla_{\vec\tau} \cdot
  (\boldsymbol\tau_i+\nabla_{\vec\tau} \chi_i)
\end{align}
on $\Sigma$.
In the above, $\boldsymbol Q^{\pm}$ denotes the value of the vector
$\boldsymbol Q$ on a prescribed side ($\pm$) of the \emph{oriented} surface
$\Sigma$, where by convention the vector $\boldsymbol\nu$ points outwards
the `$+$' side.  Equation~\eqref{eq:chi-bc} entails that the electric field
normal to the conducting sheet experiences a jump proportional to the
surface charge density on $\Sigma$. This condition also accounts for the
jump of the tangential magnetic field due to the surface current density on
$\Sigma$. Note that the term containing
$\nabla_{\vec\tau}\cdot \boldsymbol\tau_i$ in Eq.~\eqref{eq:chi-bc} can
play the role of a forcing for the Laplace equation~\eqref{eq:chi-pde}. This
term can be nonzero on a non-planar surface $\Sigma$ (e.g., a circular
nanotube).

Another boundary condition for $\chi_i$ arises from the requirement that
the electric field normal to possible edges of the conducting sheet must
vanish~\cite{MaierML18}. This condition can be written as
\begin{align}\label{eq:chi-edge}
  \boldsymbol n\cdot
  (\boldsymbol\tau_i+\nabla_{\vec\tau}\chi_i)=0\qquad \mbox{along\ edge}
\end{align}
where $\boldsymbol n$ is the outward-pointing unit vector normal to the
edge and tangential to the surface $\Sigma$. By this condition, the term
proportional to $\boldsymbol n\cdot \boldsymbol\tau_i$ can play the role of
a forcing for the Laplace equation obeyed by $\chi_i$, if the edge is
present and $\boldsymbol n\cdot \boldsymbol\tau_i\neq 0$ (e.g., for a
nanoribbon).  Equation~\eqref{eq:chi-edge} holds in the absence of a line
charge density along the edge; see the discussion and extension
in~\cite{Maier19b}.

We can provide a plausibility argument for the microscale character of
$\chi_i(\vr)$, by recalling our scaling hypothesis $\sigma\diagup d\simeq
\text{const}$. By assuming that $\chi_i(\vr)$ depends on $\vr/d$ and
non-dimensionalizing spatial coordinates, we realize that $\sigma\diagup d$
appears on the right-hand side of Eq.~\eqref{eq:chi-bc}. Thus, this
condition is independent of $d$ (as $k_0 d\to 0$).

A few remarks on the geometry with nanoribbons are in order (see
Fig.~\ref{fig:nanoribbons-rve} and Sec.~\ref{subsec:corr-phys}). In this
configuration, the only nonzero component of the corrector
$\boldsymbol\chi$ is $\chi_z$ ($\chi_x=0=\chi_y$)~\cite{Maier19b}. We can
explain the vanishing of the components $\chi_x$ and $\chi_y$ heuristically
by resorting to the above boundary value problem. Simply notice that each
of these components ($\chi_i$ for $i=x,\,y$) satisfies the Laplace equation
with \emph{homogeneous} (forcing-free) boundary conditions across the
surface $\Sigma$ and along the nanoribbon edges which form the boundary of
$\Sigma$. More precisely, $\nabla_\para\cdot \boldsymbol\tau_i=0$ on the
right-hand side of condition~\eqref{eq:chi-bc} for all $i$; and
$\boldsymbol n\cdot \boldsymbol \tau_i=0$ for $i=x,\,y$ in
Eq.~\eqref{eq:chi-edge}. Hence, by the expected uniqueness of the solution
to the cell problem for $\chi_i$, we have $\chi_x=0=\chi_y$. In contrast,
$\chi_z$ is nonzero because $\boldsymbol n\cdot \boldsymbol \tau_z=\pm 1$
along the edges. Thus, the solution to the Laplace equation for $\chi_z$ is
affected by the forcing term in the requisite boundary condition along the
edges.

Hence, for the nanoribbon setting, Eq.~\eqref{eq:eff-general} implies that
the effective permittivity tensor $\teff$ is represented by a diagonal
matrix with distinct elements, $\teff=\text{diag}(\eff_x, \eff_y, \eff_z)$
and $\eff_x\neq \eff_y\neq \eff_z\neq \eff_x$; see
Sec.~\ref{subsec:corr-phys} for more details. This observation motivates
the simplified but practically useful analysis of
Sec.~\ref{subsec:hom-Fresnel-diagonal_e}.

We note in passing that, if needed, Eq.~\eqref{eq:eff-tensor-app} combined with the above boundary value problem for $\boldsymbol \chi$ can be readily extended to a more general setting
involving tensor-valued and spatially dependent material models for
permittivitiy $\te(\omega,\boldsymbol r)$ and surface conductivity
$\underline{\sigma}(\omega, \boldsymbol r)$ \cite{Maier19b}. In this paper,
however, we restrict the discussion to the more practical situation of
homogenization effects due to the sheet geometry alone.

%%%%%%%%%%%%%%%%%%%%%%%%%%%%%%%%%%%%%%%%%%%%%%%%%%%%%%%%%%%%%%%%%%%%%%%%%%%%%%%%
%%%%%%%%%%%%%%%%%%%%%%%%%%%%%%%%%%%%%%%%%%%%%%%%%%%%%%%%%%%%%%%%%%%%%%%%%%%%%%%%
%%%%%%%%%%%%%%%%%%%%%%%%%%%%%%%%%%%%%%%%%%%%%%%%%%%%%%%%%%%%%%%%%%%%%%%%%%%%%%%%

\bibliography{references}

\begin{thebibliography}{46}%
\makeatletter
\providecommand \@ifxundefined [1]{%
 \@ifx{#1\undefined}
}%
\providecommand \@ifnum [1]{%
 \ifnum #1\expandafter \@firstoftwo
 \else \expandafter \@secondoftwo
 \fi
}%
\providecommand \@ifx [1]{%
 \ifx #1\expandafter \@firstoftwo
 \else \expandafter \@secondoftwo
 \fi
}%
\providecommand \natexlab [1]{#1}%
\providecommand \enquote  [1]{``#1''}%
\providecommand \bibnamefont  [1]{#1}%
\providecommand \bibfnamefont [1]{#1}%
\providecommand \citenamefont [1]{#1}%
\providecommand \href@noop [0]{\@secondoftwo}%
\providecommand \href [0]{\begingroup \@sanitize@url \@href}%
\providecommand \@href[1]{\@@startlink{#1}\@@href}%
\providecommand \@@href[1]{\endgroup#1\@@endlink}%
\providecommand \@sanitize@url [0]{\catcode `\\12\catcode `\$12\catcode
  `\&12\catcode `\#12\catcode `\^12\catcode `\_12\catcode `\%12\relax}%
\providecommand \@@startlink[1]{}%
\providecommand \@@endlink[0]{}%
\providecommand \url  [0]{\begingroup\@sanitize@url \@url }%
\providecommand \@url [1]{\endgroup\@href {#1}{\urlprefix }}%
\providecommand \urlprefix  [0]{URL }%
\providecommand \Eprint [0]{\href }%
\providecommand \doibase [0]{https://doi.org/}%
\providecommand \selectlanguage [0]{\@gobble}%
\providecommand \bibinfo  [0]{\@secondoftwo}%
\providecommand \bibfield  [0]{\@secondoftwo}%
\providecommand \translation [1]{[#1]}%
\providecommand \BibitemOpen [0]{}%
\providecommand \bibitemStop [0]{}%
\providecommand \bibitemNoStop [0]{.\EOS\space}%
\providecommand \EOS [0]{\spacefactor3000\relax}%
\providecommand \BibitemShut  [1]{\csname bibitem#1\endcsname}%
\let\auto@bib@innerbib\@empty
%</preamble>
\bibitem [{\citenamefont {{L.~E.~F. Foa Torres, S. Roche, and J.-C.
  Charlier}}(2014)}]{Torres2014}%
  \BibitemOpen
  \bibfield  {author} {\bibinfo {author} {\bibnamefont {{L.~E.~F. Foa Torres,
  S. Roche, and J.-C. Charlier}}},\ }\href
  {https://doi.org/10.1017/CBO9781139344364} {\emph {\bibinfo {title}
  {Introduction to Graphene-Based Nanomaterials: From Electronic Structure to
  Quantum Transport}}}\ (\bibinfo  {publisher} {Cambridge University Press},\
  \bibinfo {address} {Cambridge, UK},\ \bibinfo {year} {2014})\BibitemShut
  {NoStop}%
\bibitem [{\citenamefont {Geim}\ and\ \citenamefont
  {Grigorieva}(2013)}]{Geimetal2013}%
  \BibitemOpen
  \bibfield  {author} {\bibinfo {author} {\bibfnamefont {A.~K.}\ \bibnamefont
  {Geim}}\ and\ \bibinfo {author} {\bibfnamefont {I.~V.}\ \bibnamefont
  {Grigorieva}},\ }\bibfield  {title} {\bibinfo {title} {Van der waals
  heterostructures},\ }\href {https://doi.org/10.1038/nature12385} {\bibfield
  {journal} {\bibinfo  {journal} {Nature (London)}\ }\textbf {\bibinfo {volume}
  {499}},\ \bibinfo {pages} {419} (\bibinfo {year} {2013})}\BibitemShut
  {NoStop}%
\bibitem [{\citenamefont {{Castro Neto}}\ \emph {et~al.}(2009)\citenamefont
  {{Castro Neto}}, \citenamefont {Guinea}, \citenamefont {Peres}, \citenamefont
  {Novoselov},\ and\ \citenamefont {Geim}}]{CastroNetoetal2009}%
  \BibitemOpen
  \bibfield  {author} {\bibinfo {author} {\bibfnamefont {A.~H.}\ \bibnamefont
  {{Castro Neto}}}, \bibinfo {author} {\bibfnamefont {F.}~\bibnamefont
  {Guinea}}, \bibinfo {author} {\bibfnamefont {N.~M.~R.}\ \bibnamefont
  {Peres}}, \bibinfo {author} {\bibfnamefont {K.~S.}\ \bibnamefont
  {Novoselov}},\ and\ \bibinfo {author} {\bibfnamefont {A.~K.}\ \bibnamefont
  {Geim}},\ }\bibfield  {title} {\bibinfo {title} {The electronic properties of
  graphene},\ }\href {https://doi.org/10.1103/RevModPhys.81.109} {\bibfield
  {journal} {\bibinfo  {journal} {Rev.\ Mod.\ Phys.}\ }\textbf {\bibinfo
  {volume} {81}},\ \bibinfo {pages} {109} (\bibinfo {year} {2009})}\BibitemShut
  {NoStop}%
\bibitem [{\citenamefont {Low}\ \emph {et~al.}(2017)\citenamefont {Low},
  \citenamefont {Chaves}, \citenamefont {Caldwell}, \citenamefont {Kumar},
  \citenamefont {Fang}, \citenamefont {Avouris}, \citenamefont {Heinz},
  \citenamefont {Guinea}, \citenamefont {Mart{\'\i}n-Moreno},\ and\
  \citenamefont {Koppens}}]{Lowetal2017}%
  \BibitemOpen
  \bibfield  {author} {\bibinfo {author} {\bibfnamefont {T.}~\bibnamefont
  {Low}}, \bibinfo {author} {\bibfnamefont {A.}~\bibnamefont {Chaves}},
  \bibinfo {author} {\bibfnamefont {J.~D.}\ \bibnamefont {Caldwell}}, \bibinfo
  {author} {\bibfnamefont {A.}~\bibnamefont {Kumar}}, \bibinfo {author}
  {\bibfnamefont {N.~X.}\ \bibnamefont {Fang}}, \bibinfo {author}
  {\bibfnamefont {P.}~\bibnamefont {Avouris}}, \bibinfo {author} {\bibfnamefont
  {T.~F.}\ \bibnamefont {Heinz}}, \bibinfo {author} {\bibfnamefont
  {F.}~\bibnamefont {Guinea}}, \bibinfo {author} {\bibfnamefont
  {L.}~\bibnamefont {Mart{\'\i}n-Moreno}},\ and\ \bibinfo {author}
  {\bibfnamefont {F.}~\bibnamefont {Koppens}},\ }\bibfield  {title} {\bibinfo
  {title} {Polaritons in layered two-dimensional materials},\ }\href
  {https://doi.org/10.1038/nmat4792} {\bibfield  {journal} {\bibinfo  {journal}
  {Nat.\ Mater.}\ }\textbf {\bibinfo {volume} {16}},\ \bibinfo {pages} {182}
  (\bibinfo {year} {2017})}\BibitemShut {NoStop}%
\bibitem [{\citenamefont {Novoselov}\ \emph {et~al.}(2012)\citenamefont
  {Novoselov}, \citenamefont {Fal'ko}, \citenamefont {Colombo}, \citenamefont
  {Gellert}, \citenamefont {Schwab},\ and\ \citenamefont
  {Kim}}]{Novoselovetal2012}%
  \BibitemOpen
  \bibfield  {author} {\bibinfo {author} {\bibfnamefont {K.~S.}\ \bibnamefont
  {Novoselov}}, \bibinfo {author} {\bibfnamefont {V.~I.}\ \bibnamefont
  {Fal'ko}}, \bibinfo {author} {\bibfnamefont {L.}~\bibnamefont {Colombo}},
  \bibinfo {author} {\bibfnamefont {P.~R.}\ \bibnamefont {Gellert}}, \bibinfo
  {author} {\bibfnamefont {M.~G.}\ \bibnamefont {Schwab}},\ and\ \bibinfo
  {author} {\bibfnamefont {K.}~\bibnamefont {Kim}},\ }\bibfield  {title}
  {\bibinfo {title} {A roadmap for graphene},\ }\href
  {https://doi.org/10.1038/nature11458} {\bibfield  {journal} {\bibinfo
  {journal} {Nature (London)}\ }\textbf {\bibinfo {volume} {490}},\ \bibinfo
  {pages} {192} (\bibinfo {year} {2012})}\BibitemShut {NoStop}%
\bibitem [{\citenamefont {Li}\ \emph {et~al.}(2008)\citenamefont {Li},
  \citenamefont {Henriksen}, \citenamefont {Jiang}, \citenamefont {Hao},
  \citenamefont {Martin}, \citenamefont {Kim}, \citenamefont {Stormer},\ and\
  \citenamefont {Basov}}]{LiBasov2008}%
  \BibitemOpen
  \bibfield  {author} {\bibinfo {author} {\bibfnamefont {Z.~Q.}\ \bibnamefont
  {Li}}, \bibinfo {author} {\bibfnamefont {E.~A.}\ \bibnamefont {Henriksen}},
  \bibinfo {author} {\bibfnamefont {Z.}~\bibnamefont {Jiang}}, \bibinfo
  {author} {\bibfnamefont {Z.}~\bibnamefont {Hao}}, \bibinfo {author}
  {\bibfnamefont {M.~C.}\ \bibnamefont {Martin}}, \bibinfo {author}
  {\bibfnamefont {P.}~\bibnamefont {Kim}}, \bibinfo {author} {\bibfnamefont
  {H.~L.}\ \bibnamefont {Stormer}},\ and\ \bibinfo {author} {\bibfnamefont
  {D.~N.}\ \bibnamefont {Basov}},\ }\bibfield  {title} {\bibinfo {title} {Dirac
  charge dynamics in graphene by infrared spectroscopy},\ }\href
  {https://doi.org/10.1038/nphys989} {\bibfield  {journal} {\bibinfo  {journal}
  {Nat.\ Phys.}\ }\textbf {\bibinfo {volume} {4}},\ \bibinfo {pages} {532}
  (\bibinfo {year} {2008})}\BibitemShut {NoStop}%
\bibitem [{\citenamefont {Kim}\ \emph {et~al.}(2018)\citenamefont {Kim},
  \citenamefont {Jang}, \citenamefont {Brar}, \citenamefont {Mauser},
  \citenamefont {Kim},\ and\ \citenamefont {Atwater}}]{Atwateretal2018}%
  \BibitemOpen
  \bibfield  {author} {\bibinfo {author} {\bibfnamefont {S.}~\bibnamefont
  {Kim}}, \bibinfo {author} {\bibfnamefont {M.~S.}\ \bibnamefont {Jang}},
  \bibinfo {author} {\bibfnamefont {V.~W.}\ \bibnamefont {Brar}}, \bibinfo
  {author} {\bibfnamefont {K.~W.}\ \bibnamefont {Mauser}}, \bibinfo {author}
  {\bibfnamefont {L.}~\bibnamefont {Kim}},\ and\ \bibinfo {author}
  {\bibfnamefont {H.~A.}\ \bibnamefont {Atwater}},\ }\bibfield  {title}
  {\bibinfo {title} {Electronically tunable perfect absorption in graphene},\
  }\href {https://doi.org/10.1021/acs.nanolett.7b04393} {\bibfield  {journal}
  {\bibinfo  {journal} {Nano Lett.}\ }\textbf {\bibinfo {volume} {18}},\
  \bibinfo {pages} {971} (\bibinfo {year} {2018})}\BibitemShut {NoStop}%
\bibitem [{\citenamefont {Dai}\ \emph {et~al.}(2015)\citenamefont {Dai},
  \citenamefont {Ma}, \citenamefont {Liu}, \citenamefont {Andersen},
  \citenamefont {Fei}, \citenamefont {Goldflam}, \citenamefont {Wagner},
  \citenamefont {Watanabe}, \citenamefont {Taniguchi}, \citenamefont
  {Thiemens}, \citenamefont {Keilmann}, \citenamefont {Janssen}, \citenamefont
  {Zhu}, \citenamefont {Jarillo-Herrero}, \citenamefont {Fogler},\ and\
  \citenamefont {Basov}}]{Daietal2015}%
  \BibitemOpen
  \bibfield  {author} {\bibinfo {author} {\bibfnamefont {S.}~\bibnamefont
  {Dai}}, \bibinfo {author} {\bibfnamefont {Q.}~\bibnamefont {Ma}}, \bibinfo
  {author} {\bibfnamefont {M.~K.}\ \bibnamefont {Liu}}, \bibinfo {author}
  {\bibfnamefont {T.}~\bibnamefont {Andersen}}, \bibinfo {author}
  {\bibfnamefont {Z.}~\bibnamefont {Fei}}, \bibinfo {author} {\bibfnamefont
  {M.~D.}\ \bibnamefont {Goldflam}}, \bibinfo {author} {\bibfnamefont
  {M.}~\bibnamefont {Wagner}}, \bibinfo {author} {\bibfnamefont
  {K.}~\bibnamefont {Watanabe}}, \bibinfo {author} {\bibfnamefont
  {T.}~\bibnamefont {Taniguchi}}, \bibinfo {author} {\bibfnamefont
  {M.}~\bibnamefont {Thiemens}}, \bibinfo {author} {\bibfnamefont
  {F.}~\bibnamefont {Keilmann}}, \bibinfo {author} {\bibfnamefont {G.~C.
  A.~M.}\ \bibnamefont {Janssen}}, \bibinfo {author} {\bibfnamefont {S.-E.}\
  \bibnamefont {Zhu}}, \bibinfo {author} {\bibfnamefont {P.}~\bibnamefont
  {Jarillo-Herrero}}, \bibinfo {author} {\bibfnamefont {M.~M.}\ \bibnamefont
  {Fogler}},\ and\ \bibinfo {author} {\bibfnamefont {D.~N.}\ \bibnamefont
  {Basov}},\ }\bibfield  {title} {\bibinfo {title} {Graphene on hexagonal boron
  nitride as a tunable hyperbolic metamaterial},\ }\href
  {https://doi.org/10.1038/nnano.2015.131} {\bibfield  {journal} {\bibinfo
  {journal} {Nat.\ Nano}\ }\textbf {\bibinfo {volume} {10}},\ \bibinfo {pages}
  {682} (\bibinfo {year} {2015})}\BibitemShut {NoStop}%
\bibitem [{\citenamefont {Nemilentsau}\ \emph {et~al.}(2016)\citenamefont
  {Nemilentsau}, \citenamefont {Low},\ and\ \citenamefont
  {Hanson}}]{Nemilentsau2016}%
  \BibitemOpen
  \bibfield  {author} {\bibinfo {author} {\bibfnamefont {A.}~\bibnamefont
  {Nemilentsau}}, \bibinfo {author} {\bibfnamefont {T.}~\bibnamefont {Low}},\
  and\ \bibinfo {author} {\bibfnamefont {G.}~\bibnamefont {Hanson}},\
  }\bibfield  {title} {\bibinfo {title} {Anisotropic 2d materials for tunable
  hyperbolic plasmonics},\ }\href
  {https://doi.org/10.1103/PhysRevLett.116.066804} {\bibfield  {journal}
  {\bibinfo  {journal} {Phys.\ Rev.\ Lett.}\ }\textbf {\bibinfo {volume}
  {116}},\ \bibinfo {pages} {066804} (\bibinfo {year} {2016})}\BibitemShut
  {NoStop}%
\bibitem [{\citenamefont {Mahmoodi}\ \emph {et~al.}(2019)\citenamefont
  {Mahmoodi}, \citenamefont {Tavassoli}, \citenamefont {Takayama},
  \citenamefont {Sukham}, \citenamefont {Malureanu},\ and\ \citenamefont
  {Lavrinenko}}]{Mahmoodi2019}%
  \BibitemOpen
  \bibfield  {author} {\bibinfo {author} {\bibfnamefont {M.}~\bibnamefont
  {Mahmoodi}}, \bibinfo {author} {\bibfnamefont {S.~H.}\ \bibnamefont
  {Tavassoli}}, \bibinfo {author} {\bibfnamefont {O.}~\bibnamefont {Takayama}},
  \bibinfo {author} {\bibfnamefont {J.}~\bibnamefont {Sukham}}, \bibinfo
  {author} {\bibfnamefont {R.}~\bibnamefont {Malureanu}},\ and\ \bibinfo
  {author} {\bibfnamefont {A.~V.}\ \bibnamefont {Lavrinenko}},\ }\bibfield
  {title} {\bibinfo {title} {Existence conditions of high-k modes in finite
  hyperbolic metamaterials},\ }\href {https://doi.org/10.1002/lpor.201800253}
  {\bibfield  {journal} {\bibinfo  {journal} {Laser Photon.\ Rev.}\ }\textbf
  {\bibinfo {volume} {13}},\ \bibinfo {pages} {1800253} (\bibinfo {year}
  {2019})}\BibitemShut {NoStop}%
\bibitem [{\citenamefont {Deng}\ \emph {et~al.}(2018)\citenamefont {Deng},
  \citenamefont {Chen}, \citenamefont {Huang},\ and\ \citenamefont
  {Ye}}]{Dengetal2018}%
  \BibitemOpen
  \bibfield  {author} {\bibinfo {author} {\bibfnamefont {H.}~\bibnamefont
  {Deng}}, \bibinfo {author} {\bibfnamefont {Y.}~\bibnamefont {Chen}}, \bibinfo
  {author} {\bibfnamefont {C.}~\bibnamefont {Huang}},\ and\ \bibinfo {author}
  {\bibfnamefont {F.}~\bibnamefont {Ye}},\ }\bibfield  {title} {\bibinfo
  {title} {Topological interface modes in photonic superlattices containing
  negative-index materials},\ }\href
  {https://doi.org/10.1209/0295-5075/124/64001} {\bibfield  {journal} {\bibinfo
   {journal} {Europhys.\ Lett.}\ }\textbf {\bibinfo {volume} {124}},\ \bibinfo
  {pages} {64001} (\bibinfo {year} {2018})}\BibitemShut {NoStop}%
\bibitem [{\citenamefont {Deng}\ \emph {et~al.}(2015)\citenamefont {Deng},
  \citenamefont {Ye}, \citenamefont {Malomed}, \citenamefont {Chen},\ and\
  \citenamefont {Panoiu}}]{Dengetal2015}%
  \BibitemOpen
  \bibfield  {author} {\bibinfo {author} {\bibfnamefont {H.}~\bibnamefont
  {Deng}}, \bibinfo {author} {\bibfnamefont {F.}~\bibnamefont {Ye}}, \bibinfo
  {author} {\bibfnamefont {B.~A.}\ \bibnamefont {Malomed}}, \bibinfo {author}
  {\bibfnamefont {X.}~\bibnamefont {Chen}},\ and\ \bibinfo {author}
  {\bibfnamefont {N.~C.}\ \bibnamefont {Panoiu}},\ }\bibfield  {title}
  {\bibinfo {title} {Optically and electrically tunable dirac points and
  {Z}itterbewegung in graphene-based photonic superlattices},\ }\href
  {https://doi.org/10.1103/PhysRevB.91.201402} {\bibfield  {journal} {\bibinfo
  {journal} {Phys.\ Rev. B}\ }\textbf {\bibinfo {volume} {91}},\ \bibinfo
  {pages} {201402(R)} (\bibinfo {year} {2015})}\BibitemShut {NoStop}%
\bibitem [{\citenamefont {Zhukovsky}\ \emph {et~al.}(2014)\citenamefont
  {Zhukovsky}, \citenamefont {Andryieuski}, \citenamefont {Sipe},\ and\
  \citenamefont {Lavrinenko}}]{Zhukovsky2014}%
  \BibitemOpen
  \bibfield  {author} {\bibinfo {author} {\bibfnamefont {S.~V.}\ \bibnamefont
  {Zhukovsky}}, \bibinfo {author} {\bibfnamefont {A.}~\bibnamefont
  {Andryieuski}}, \bibinfo {author} {\bibfnamefont {J.~E.}\ \bibnamefont
  {Sipe}},\ and\ \bibinfo {author} {\bibfnamefont {A.~V.}\ \bibnamefont
  {Lavrinenko}},\ }\bibfield  {title} {\bibinfo {title} {From surface to volume
  plasmons in hyperbolic metamaterials: General existence conditions for bulk
  {high-$\mathbf k$} waves in metal-dielectric and graphene-dielectric
  multilayers},\ }\href {https://doi.org/10.1103/PhysRevB.90.155429} {\bibfield
   {journal} {\bibinfo  {journal} {Phys.\ Rev. B}\ }\textbf {\bibinfo {volume}
  {90}},\ \bibinfo {pages} {2155429} (\bibinfo {year} {2014})}\BibitemShut
  {NoStop}%
\bibitem [{\citenamefont {Maier}\ \emph
  {et~al.}(2018{\natexlab{a}})\citenamefont {Maier}, \citenamefont
  {Mattheakis}, \citenamefont {Kaxiras}, \citenamefont {Luskin},\ and\
  \citenamefont {Margetis}}]{Maier18}%
  \BibitemOpen
  \bibfield  {author} {\bibinfo {author} {\bibfnamefont {M.}~\bibnamefont
  {Maier}}, \bibinfo {author} {\bibfnamefont {M.}~\bibnamefont {Mattheakis}},
  \bibinfo {author} {\bibfnamefont {E.}~\bibnamefont {Kaxiras}}, \bibinfo
  {author} {\bibfnamefont {M.}~\bibnamefont {Luskin}},\ and\ \bibinfo {author}
  {\bibfnamefont {D.}~\bibnamefont {Margetis}},\ }\bibfield  {title} {\bibinfo
  {title} {Universal behavior of dispersive {D}irac cone in gradient-index
  plasmonic metamaterials},\ }\href
  {https://doi.org/10.1103/PhysRevB.97.035307} {\bibfield  {journal} {\bibinfo
  {journal} {Phys.\ Rev. B}\ }\textbf {\bibinfo {volume} {97}},\ \bibinfo
  {pages} {035307} (\bibinfo {year} {2018}{\natexlab{a}})}\BibitemShut
  {NoStop}%
\bibitem [{\citenamefont {Choy}(1999)}]{Choy1999}%
  \BibitemOpen
  \bibfield  {author} {\bibinfo {author} {\bibfnamefont {T.~C.}\ \bibnamefont
  {Choy}},\ }\href@noop {} {\emph {\bibinfo {title} {Effective Medium Theory:
  Principles and Applications}}}\ (\bibinfo  {publisher} {Clarendon Press},\
  \bibinfo {address} {Oxford, UK},\ \bibinfo {year} {1999})\ \bibinfo {note}
  {chap. 3}\BibitemShut {NoStop}%
\bibitem [{\citenamefont {Pavliotis}\ and\ \citenamefont
  {Stuart}(2007)}]{StuartPavliotis2007}%
  \BibitemOpen
  \bibfield  {author} {\bibinfo {author} {\bibfnamefont {G.}~\bibnamefont
  {Pavliotis}}\ and\ \bibinfo {author} {\bibfnamefont {A.~M.}\ \bibnamefont
  {Stuart}},\ }\href@noop {} {\emph {\bibinfo {title} {Multiscale methods:
  Averaging and homogenization}}}\ (\bibinfo  {publisher} {Springer},\ \bibinfo
  {address} {Berlin, Germany},\ \bibinfo {year} {2007})\BibitemShut {NoStop}%
\bibitem [{\citenamefont {Maier}\ \emph {et~al.}(2019)\citenamefont {Maier},
  \citenamefont {Mattheakis}, \citenamefont {Kaxiras}, \citenamefont {Luskin},\
  and\ \citenamefont {Margetis}}]{Maier19b}%
  \BibitemOpen
  \bibfield  {author} {\bibinfo {author} {\bibfnamefont {M.}~\bibnamefont
  {Maier}}, \bibinfo {author} {\bibfnamefont {M.}~\bibnamefont {Mattheakis}},
  \bibinfo {author} {\bibfnamefont {E.}~\bibnamefont {Kaxiras}}, \bibinfo
  {author} {\bibfnamefont {M.}~\bibnamefont {Luskin}},\ and\ \bibinfo {author}
  {\bibfnamefont {D.}~\bibnamefont {Margetis}},\ }\bibfield  {title} {\bibinfo
  {title} {Homogenization of plasmonic crystals: seeking the epsilon-near-zero
  effect},\ }\href {https://doi.org/10.1098/rspa.2019.0220} {\bibfield
  {journal} {\bibinfo  {journal} {Proc.\ R.\ Soc.\ A}\ }\textbf {\bibinfo
  {volume} {475}},\ \bibinfo {pages} {20190220} (\bibinfo {year}
  {2019})}\BibitemShut {NoStop}%
\bibitem [{\citenamefont {Yan}\ \emph {et~al.}(2012)\citenamefont {Yan},
  \citenamefont {Li}, \citenamefont {Chandra}, \citenamefont {Tulevski},
  \citenamefont {Wu}, \citenamefont {Freitag}, \citenamefont {Zhu},
  \citenamefont {Avouris},\ and\ \citenamefont {Xia}}]{Yanetal2012}%
  \BibitemOpen
  \bibfield  {author} {\bibinfo {author} {\bibfnamefont {H.}~\bibnamefont
  {Yan}}, \bibinfo {author} {\bibfnamefont {X.}~\bibnamefont {Li}}, \bibinfo
  {author} {\bibfnamefont {B.}~\bibnamefont {Chandra}}, \bibinfo {author}
  {\bibfnamefont {G.}~\bibnamefont {Tulevski}}, \bibinfo {author}
  {\bibfnamefont {Y.}~\bibnamefont {Wu}}, \bibinfo {author} {\bibfnamefont
  {M.}~\bibnamefont {Freitag}}, \bibinfo {author} {\bibfnamefont
  {W.}~\bibnamefont {Zhu}}, \bibinfo {author} {\bibfnamefont {P.}~\bibnamefont
  {Avouris}},\ and\ \bibinfo {author} {\bibfnamefont {F.}~\bibnamefont {Xia}},\
  }\bibfield  {title} {\bibinfo {title} {Tunable infrared plasmonic devices
  using graphene/insulator stacks},\ }\href
  {https://doi.org/10.1038/NNANO.2012.59} {\bibfield  {journal} {\bibinfo
  {journal} {Nat.\ Nano}\ }\textbf {\bibinfo {volume} {7}},\ \bibinfo {pages}
  {330} (\bibinfo {year} {2012})}\BibitemShut {NoStop}%
\bibitem [{\citenamefont {Yao}\ \emph {et~al.}(2018)\citenamefont {Yao},
  \citenamefont {Liu}, \citenamefont {Huang}, \citenamefont {Choi1},
  \citenamefont {Xie}, \citenamefont {Flores}, \citenamefont {Wu},
  \citenamefont {Yu}, \citenamefont {Kwong}, \citenamefont {Huang},
  \citenamefont {Rao}, \citenamefont {Duan},\ and\ \citenamefont
  {Wong}}]{Yaoetal2018}%
  \BibitemOpen
  \bibfield  {author} {\bibinfo {author} {\bibfnamefont {B.}~\bibnamefont
  {Yao}}, \bibinfo {author} {\bibfnamefont {Y.}~\bibnamefont {Liu}}, \bibinfo
  {author} {\bibfnamefont {S.-W.}\ \bibnamefont {Huang}}, \bibinfo {author}
  {\bibfnamefont {C.}~\bibnamefont {Choi1}}, \bibinfo {author} {\bibfnamefont
  {Z.}~\bibnamefont {Xie}}, \bibinfo {author} {\bibfnamefont {J.~F.}\
  \bibnamefont {Flores}}, \bibinfo {author} {\bibfnamefont {Y.}~\bibnamefont
  {Wu}}, \bibinfo {author} {\bibfnamefont {M.}~\bibnamefont {Yu}}, \bibinfo
  {author} {\bibfnamefont {D.-L.}\ \bibnamefont {Kwong}}, \bibinfo {author}
  {\bibfnamefont {Y.}~\bibnamefont {Huang}}, \bibinfo {author} {\bibfnamefont
  {Y.}~\bibnamefont {Rao}}, \bibinfo {author} {\bibfnamefont {X.}~\bibnamefont
  {Duan}},\ and\ \bibinfo {author} {\bibfnamefont {C.~W.}\ \bibnamefont
  {Wong}},\ }\bibfield  {title} {\bibinfo {title} {Broadband gate-tunable
  terahertz plasmons in graphene heterostructures},\ }\href
  {https://doi.org/10.1038/s41566-017-0054-7} {\bibfield  {journal} {\bibinfo
  {journal} {Nat.\ Photon.}\ }\textbf {\bibinfo {volume} {12}},\ \bibinfo
  {pages} {22} (\bibinfo {year} {2018})}\BibitemShut {NoStop}%
\bibitem [{\citenamefont {Ma}\ \emph {et~al.}(2019)\citenamefont {Ma},
  \citenamefont {Salamin}, \citenamefont {Baeuerle}, \citenamefont {Josten},
  \citenamefont {Heni}, \citenamefont {Emboras},\ and\ \citenamefont
  {Leuthold}}]{Ma2019}%
  \BibitemOpen
  \bibfield  {author} {\bibinfo {author} {\bibfnamefont {P.}~\bibnamefont
  {Ma}}, \bibinfo {author} {\bibfnamefont {Y.}~\bibnamefont {Salamin}},
  \bibinfo {author} {\bibfnamefont {B.}~\bibnamefont {Baeuerle}}, \bibinfo
  {author} {\bibfnamefont {A.}~\bibnamefont {Josten}}, \bibinfo {author}
  {\bibfnamefont {W.}~\bibnamefont {Heni}}, \bibinfo {author} {\bibfnamefont
  {A.}~\bibnamefont {Emboras}},\ and\ \bibinfo {author} {\bibfnamefont
  {J.}~\bibnamefont {Leuthold}},\ }\bibfield  {title} {\bibinfo {title}
  {Plasmonically enhanced graphene photodetector featuring 100 gbit/s data
  reception, high responsivity, and compact size},\ }\href
  {https://doi.org/10.1021/acsphotonics.8b01234} {\bibfield  {journal}
  {\bibinfo  {journal} {ACS Photon.}\ }\textbf {\bibinfo {volume} {6}},\
  \bibinfo {pages} {154} (\bibinfo {year} {2019})}\BibitemShut {NoStop}%
\bibitem [{\citenamefont {Nematpour}\ \emph {et~al.}(2019)\citenamefont
  {Nematpour}, \citenamefont {Lim}, \citenamefont {Piegari}, \citenamefont
  {Lancellotti}, \citenamefont {Hu},\ and\ \citenamefont
  {Grilli}}]{Nematpour2019}%
  \BibitemOpen
  \bibfield  {author} {\bibinfo {author} {\bibfnamefont {A.}~\bibnamefont
  {Nematpour}}, \bibinfo {author} {\bibfnamefont {N.}~\bibnamefont {Lim}},
  \bibinfo {author} {\bibfnamefont {A.}~\bibnamefont {Piegari}}, \bibinfo
  {author} {\bibfnamefont {L.}~\bibnamefont {Lancellotti}}, \bibinfo {author}
  {\bibfnamefont {G.}~\bibnamefont {Hu}},\ and\ \bibinfo {author}
  {\bibfnamefont {M.~L.}\ \bibnamefont {Grilli}},\ }\bibfield  {title}
  {\bibinfo {title} {Experimental near infrared absorption enhancement of
  graphene layers in an optical resonant cavity},\ }\href
  {https://doi.org/10.1088/1361-6528/ab346d} {\bibfield  {journal} {\bibinfo
  {journal} {Nanotechnology}\ }\textbf {\bibinfo {volume} {30}},\ \bibinfo
  {pages} {445201} (\bibinfo {year} {2019})}\BibitemShut {NoStop}%
\bibitem [{\citenamefont {Kumar}\ \emph {et~al.}(2015)\citenamefont {Kumar},
  \citenamefont {Low}, \citenamefont {Fung}, \citenamefont {Avouris},\ and\
  \citenamefont {Fang}}]{KumarAvourisetal2015}%
  \BibitemOpen
  \bibfield  {author} {\bibinfo {author} {\bibfnamefont {A.}~\bibnamefont
  {Kumar}}, \bibinfo {author} {\bibfnamefont {T.}~\bibnamefont {Low}}, \bibinfo
  {author} {\bibfnamefont {K.~H.}\ \bibnamefont {Fung}}, \bibinfo {author}
  {\bibfnamefont {P.}~\bibnamefont {Avouris}},\ and\ \bibinfo {author}
  {\bibfnamefont {N.~X.}\ \bibnamefont {Fang}},\ }\bibfield  {title} {\bibinfo
  {title} {Tunable light-matter interaction and the role of hyperbolicity in
  graphene-hbn system},\ }\href {https://doi.org/10.1021/acs.nanolett.5b01191}
  {\bibfield  {journal} {\bibinfo  {journal} {Nano Lett.}\ }\textbf {\bibinfo
  {volume} {15}},\ \bibinfo {pages} {3172} (\bibinfo {year}
  {2015})}\BibitemShut {NoStop}%
\bibitem [{\citenamefont {Hu}\ \emph {et~al.}(2019)\citenamefont {Hu},
  \citenamefont {Yang}, \citenamefont {Guo}, \citenamefont {Khaliji},
  \citenamefont {Biswas}, \citenamefont {de~Abajo}, \citenamefont {Low},
  \citenamefont {Sun},\ and\ \citenamefont {Dai}}]{HuAvourisetal2019}%
  \BibitemOpen
  \bibfield  {author} {\bibinfo {author} {\bibfnamefont {H.}~\bibnamefont
  {Hu}}, \bibinfo {author} {\bibfnamefont {X.}~\bibnamefont {Yang}}, \bibinfo
  {author} {\bibfnamefont {X.}~\bibnamefont {Guo}}, \bibinfo {author}
  {\bibfnamefont {K.}~\bibnamefont {Khaliji}}, \bibinfo {author} {\bibfnamefont
  {S.~R.}\ \bibnamefont {Biswas}}, \bibinfo {author} {\bibfnamefont {F.~J.~G.}\
  \bibnamefont {de~Abajo}}, \bibinfo {author} {\bibfnamefont {T.}~\bibnamefont
  {Low}}, \bibinfo {author} {\bibfnamefont {Z.}~\bibnamefont {Sun}},\ and\
  \bibinfo {author} {\bibfnamefont {Q.}~\bibnamefont {Dai}},\ }\bibfield
  {title} {\bibinfo {title} {Gas identification with graphene plasmons},\
  }\href {https://doi.org/10.1038/s41467-019-09008-0} {\bibfield  {journal}
  {\bibinfo  {journal} {Nat.\ Commun.}\ }\textbf {\bibinfo {volume} {10}},\
  \bibinfo {pages} {1131} (\bibinfo {year} {2019})}\BibitemShut {NoStop}%
\bibitem [{\citenamefont {Lee}\ \emph {et~al.}(2019)\citenamefont {Lee},
  \citenamefont {Yoo}, \citenamefont {Avouris}, \citenamefont {Low},\ and\
  \citenamefont {Oh}}]{LeeAvourisetal2019}%
  \BibitemOpen
  \bibfield  {author} {\bibinfo {author} {\bibfnamefont {I.-H.}\ \bibnamefont
  {Lee}}, \bibinfo {author} {\bibfnamefont {D.}~\bibnamefont {Yoo}}, \bibinfo
  {author} {\bibfnamefont {P.}~\bibnamefont {Avouris}}, \bibinfo {author}
  {\bibfnamefont {T.}~\bibnamefont {Low}},\ and\ \bibinfo {author}
  {\bibfnamefont {S.-H.}\ \bibnamefont {Oh}},\ }\bibfield  {title} {\bibinfo
  {title} {Graphene acoustic plasmon resonator for ultrasensitive infrared
  spectroscopy},\ }\href {https://doi.org/10.1038/s41565-019-0363-8} {\bibfield
   {journal} {\bibinfo  {journal} {Nat.\ Nano}\ }\textbf {\bibinfo {volume}
  {14}},\ \bibinfo {pages} {313} (\bibinfo {year} {2019})}\BibitemShut
  {NoStop}%
\bibitem [{\citenamefont {Si}\ and\ \citenamefont {Sun}(2017)}]{SiSun2017}%
  \BibitemOpen
  \bibfield  {author} {\bibinfo {author} {\bibfnamefont {J.}~\bibnamefont
  {Si}}\ and\ \bibinfo {author} {\bibfnamefont {C.}~\bibnamefont {Sun}},\
  }\bibfield  {title} {\bibinfo {title} {On the optical performance of
  composite structures of graphene and photonic crystals at infrared
  wavelengths},\ }\href {https://doi.org/10.1063/1.4998478} {\bibfield
  {journal} {\bibinfo  {journal} {J.\ Appl.\ Phys.}\ }\textbf {\bibinfo
  {volume} {122}},\ \bibinfo {pages} {133104} (\bibinfo {year}
  {2017})}\BibitemShut {NoStop}%
\bibitem [{\citenamefont {Hu}\ \emph {et~al.}(2018)\citenamefont {Hu},
  \citenamefont {Guo}, \citenamefont {Hu}, \citenamefont {Sun}, \citenamefont
  {Yang},\ and\ \citenamefont {Dai}}]{Huetal2018}%
  \BibitemOpen
  \bibfield  {author} {\bibinfo {author} {\bibfnamefont {H.}~\bibnamefont
  {Hu}}, \bibinfo {author} {\bibfnamefont {X.}~\bibnamefont {Guo}}, \bibinfo
  {author} {\bibfnamefont {D.}~\bibnamefont {Hu}}, \bibinfo {author}
  {\bibfnamefont {Z.}~\bibnamefont {Sun}}, \bibinfo {author} {\bibfnamefont
  {X.}~\bibnamefont {Yang}},\ and\ \bibinfo {author} {\bibfnamefont
  {Q.}~\bibnamefont {Dai}},\ }\bibfield  {title} {\bibinfo {title} {Flexible
  and electrically tunable plasmons in graphene-- mica heterostructures},\
  }\href {https://doi.org/10.1002/advs.201800175} {\bibfield  {journal}
  {\bibinfo  {journal} {Adv.\ Sci.}\ }\textbf {\bibinfo {volume} {5}},\
  \bibinfo {pages} {1800175} (\bibinfo {year} {2018})}\BibitemShut {NoStop}%
\bibitem [{\citenamefont {Kim}\ \emph {et~al.}(2019)\citenamefont {Kim},
  \citenamefont {Menabde}, \citenamefont {Brar},\ and\ \citenamefont
  {Jang}}]{Kimetal2019}%
  \BibitemOpen
  \bibfield  {author} {\bibinfo {author} {\bibfnamefont {S.}~\bibnamefont
  {Kim}}, \bibinfo {author} {\bibfnamefont {S.~G.}\ \bibnamefont {Menabde}},
  \bibinfo {author} {\bibfnamefont {V.~W.}\ \bibnamefont {Brar}},\ and\
  \bibinfo {author} {\bibfnamefont {M.~S.}\ \bibnamefont {Jang}},\ }\bibfield
  {title} {\bibinfo {title} {Functional mid-infrared polaritonics in van der
  waals crystals},\ }\href {https://doi.org/10.1002/adom.201901194} {\bibfield
  {journal} {\bibinfo  {journal} {Adv.\ Opt.\ Mater.}\ ,\ \bibinfo {pages}
  {1901194}} (\bibinfo {year} {2019})}\BibitemShut {NoStop}%
\bibitem [{\citenamefont {Yeh}(2005)}]{Yeh2005}%
  \BibitemOpen
  \bibfield  {author} {\bibinfo {author} {\bibfnamefont {P.}~\bibnamefont
  {Yeh}},\ }\href@noop {} {\emph {\bibinfo {title} {Optical Waves in Layered
  Media}}},\ \bibinfo {edition} {2nd}\ ed.\ (\bibinfo  {publisher} {Wiley},\
  \bibinfo {address} {Hoboken, NJ},\ \bibinfo {year} {2005})\ \bibinfo {note}
  {chap. 5}\BibitemShut {NoStop}%
\bibitem [{\citenamefont {Haus}(1984)}]{Haus1984}%
  \BibitemOpen
  \bibfield  {author} {\bibinfo {author} {\bibfnamefont {H.~A.}\ \bibnamefont
  {Haus}},\ }\href@noop {} {\emph {\bibinfo {title} {Waves and Fields in
  Optoelectronics}}}\ (\bibinfo  {publisher} {Prentice Hall},\ \bibinfo
  {address} {Englewood Cliffs, NJ},\ \bibinfo {year} {1984})\ \bibinfo {note}
  {chap. 5}\BibitemShut {NoStop}%
\bibitem [{\citenamefont {Mattheakis}\ \emph {et~al.}(2016)\citenamefont
  {Mattheakis}, \citenamefont {Valagiannopoulos},\ and\ \citenamefont
  {Kaxiras}}]{Mattheakisetal2016}%
  \BibitemOpen
  \bibfield  {author} {\bibinfo {author} {\bibfnamefont {M.}~\bibnamefont
  {Mattheakis}}, \bibinfo {author} {\bibfnamefont {C.~A.}\ \bibnamefont
  {Valagiannopoulos}},\ and\ \bibinfo {author} {\bibfnamefont {E.}~\bibnamefont
  {Kaxiras}},\ }\bibfield  {title} {\bibinfo {title} {Epsilon-near-zero
  behavior from plasmonic dirac point: Theory and realization using
  two-dimensional materials},\ }\href
  {https://doi.org/10.1103/PhysRevB.94.201404} {\bibfield  {journal} {\bibinfo
  {journal} {Phys.\ Rev.\ B}\ }\textbf {\bibinfo {volume} {94}},\ \bibinfo
  {pages} {201404(R)} (\bibinfo {year} {2016})}\BibitemShut {NoStop}%
\bibitem [{\citenamefont {Maier}\ \emph {et~al.}(2017)\citenamefont {Maier},
  \citenamefont {Margetis},\ and\ \citenamefont {Luskin}}]{Maier17}%
  \BibitemOpen
  \bibfield  {author} {\bibinfo {author} {\bibfnamefont {M.}~\bibnamefont
  {Maier}}, \bibinfo {author} {\bibfnamefont {D.}~\bibnamefont {Margetis}},\
  and\ \bibinfo {author} {\bibfnamefont {M.}~\bibnamefont {Luskin}},\
  }\bibfield  {title} {\bibinfo {title} {Dipole excitation of surface plasmon
  on a conducting sheet: finite element approximation and validation},\ }\href
  {https://doi.org/10.1016/j.jcp.2017.03.014} {\bibfield  {journal} {\bibinfo
  {journal} {J.\ Comp.\ Phys.}\ }\textbf {\bibinfo {volume} {339}},\ \bibinfo
  {pages} {126} (\bibinfo {year} {2017})}\BibitemShut {NoStop}%
\bibitem [{\citenamefont {Silveirinha}\ and\ \citenamefont
  {Engheta}(2006)}]{SilveirinhaEngheta2006}%
  \BibitemOpen
  \bibfield  {author} {\bibinfo {author} {\bibfnamefont {M.}~\bibnamefont
  {Silveirinha}}\ and\ \bibinfo {author} {\bibfnamefont {N.}~\bibnamefont
  {Engheta}},\ }\bibfield  {title} {\bibinfo {title} {Tunneling of
  electromagnetic energy through subwavelength channels and bends using
  $\epsilon$-near-zero materials},\ }\href
  {https://doi.org/10.1103/PhysRevLett.97.157403} {\bibfield  {journal}
  {\bibinfo  {journal} {Phys.\ Rev.\ Lett.}\ }\textbf {\bibinfo {volume}
  {97}},\ \bibinfo {pages} {157403} (\bibinfo {year} {2006})}\BibitemShut
  {NoStop}%
\bibitem [{\citenamefont {Huang}\ \emph {et~al.}(2011)\citenamefont {Huang},
  \citenamefont {Lai}, \citenamefont {Hang}, \citenamefont {Zheng},\ and\
  \citenamefont {Chan}}]{Huangetal2011}%
  \BibitemOpen
  \bibfield  {author} {\bibinfo {author} {\bibfnamefont {X.}~\bibnamefont
  {Huang}}, \bibinfo {author} {\bibfnamefont {Y.}~\bibnamefont {Lai}}, \bibinfo
  {author} {\bibfnamefont {Z.~H.}\ \bibnamefont {Hang}}, \bibinfo {author}
  {\bibfnamefont {H.}~\bibnamefont {Zheng}},\ and\ \bibinfo {author}
  {\bibfnamefont {C.~T.}\ \bibnamefont {Chan}},\ }\bibfield  {title} {\bibinfo
  {title} {Dirac cones induced by accidental degeneracy in photonic crystals
  and zero-refractive-index materials},\ }\href
  {https://doi.org/10.1038/nmat3030} {\bibfield  {journal} {\bibinfo  {journal}
  {Nat.\ Mater.}\ }\textbf {\bibinfo {volume} {10}},\ \bibinfo {pages} {582}
  (\bibinfo {year} {2011})}\BibitemShut {NoStop}%
\bibitem [{\citenamefont {Moitra}\ \emph {et~al.}(2013)\citenamefont {Moitra},
  \citenamefont {Yang}, \citenamefont {Anderson}, \citenamefont {Kravchenko},
  \citenamefont {Briggs},\ and\ \citenamefont {Valentine}}]{Moitraetal2013}%
  \BibitemOpen
  \bibfield  {author} {\bibinfo {author} {\bibfnamefont {P.}~\bibnamefont
  {Moitra}}, \bibinfo {author} {\bibfnamefont {Y.}~\bibnamefont {Yang}},
  \bibinfo {author} {\bibfnamefont {Z.}~\bibnamefont {Anderson}}, \bibinfo
  {author} {\bibfnamefont {I.~I.}\ \bibnamefont {Kravchenko}}, \bibinfo
  {author} {\bibfnamefont {D.~P.}\ \bibnamefont {Briggs}},\ and\ \bibinfo
  {author} {\bibfnamefont {J.}~\bibnamefont {Valentine}},\ }\bibfield  {title}
  {\bibinfo {title} {Realization of an all-dielectric zero-index optical
  metamaterial},\ }\href {https://doi.org/10.1038/nphoton.2013.214} {\bibfield
  {journal} {\bibinfo  {journal} {Nat.\ Photon.}\ }\textbf {\bibinfo {volume}
  {7}},\ \bibinfo {pages} {791} (\bibinfo {year} {2013})}\BibitemShut {NoStop}%
\bibitem [{\citenamefont {Li}\ \emph {et~al.}(2015)\citenamefont {Li},
  \citenamefont {Kita}, \citenamefont {{Mu{\~n}oz}}, \citenamefont {Reshef},
  \citenamefont {Vulis}, \citenamefont {Yin}, \citenamefont {Lon{\v c}ar},\
  and\ \citenamefont {Mazur}}]{LiMazuretal2015}%
  \BibitemOpen
  \bibfield  {author} {\bibinfo {author} {\bibfnamefont {Y.}~\bibnamefont
  {Li}}, \bibinfo {author} {\bibfnamefont {S.}~\bibnamefont {Kita}}, \bibinfo
  {author} {\bibfnamefont {P.}~\bibnamefont {{Mu{\~n}oz}}}, \bibinfo {author}
  {\bibfnamefont {O.}~\bibnamefont {Reshef}}, \bibinfo {author} {\bibfnamefont
  {D.~I.}\ \bibnamefont {Vulis}}, \bibinfo {author} {\bibfnamefont
  {M.}~\bibnamefont {Yin}}, \bibinfo {author} {\bibfnamefont {M.}~\bibnamefont
  {Lon{\v c}ar}},\ and\ \bibinfo {author} {\bibfnamefont {E.}~\bibnamefont
  {Mazur}},\ }\bibfield  {title} {\bibinfo {title} {On-chip zero-index
  metamaterials},\ }\href {https://doi.org/10.1038/nphoton.2015.198} {\bibfield
   {journal} {\bibinfo  {journal} {Nat.\ Photon.}\ }\textbf {\bibinfo {volume}
  {9}},\ \bibinfo {pages} {738} (\bibinfo {year} {2015})}\BibitemShut {NoStop}%
\bibitem [{\citenamefont {Mattheakis}\ \emph {et~al.}(2019)\citenamefont
  {Mattheakis}, \citenamefont {Maier}, \citenamefont {Boo},\ and\ \citenamefont
  {Kaxiras}}]{Maier19a}%
  \BibitemOpen
  \bibfield  {author} {\bibinfo {author} {\bibfnamefont {M.}~\bibnamefont
  {Mattheakis}}, \bibinfo {author} {\bibfnamefont {M.}~\bibnamefont {Maier}},
  \bibinfo {author} {\bibfnamefont {W.~X.}\ \bibnamefont {Boo}},\ and\ \bibinfo
  {author} {\bibfnamefont {E.}~\bibnamefont {Kaxiras}},\ }\bibfield  {title}
  {\bibinfo {title} {Graphene epsilon-near-zero plasmonic crystals},\ }in\
  \href {https://doi.org/10.1145/3345312.3345496} {\emph {\bibinfo {booktitle}
  {Proceedings of the Sixth Annual ACM International Conference on Nanoscale
  Computing and Communication}}},\ \bibinfo {series and number} {NANOCOM '19}\
  (\bibinfo {year} {2019})\ pp.\ \bibinfo {pages} {2:1--2:6}\BibitemShut
  {NoStop}%
\bibitem [{\citenamefont {{A.~M Urbas et al.}}(2016)}]{Urbas2016}%
  \BibitemOpen
  \bibfield  {author} {\bibinfo {author} {\bibnamefont {{A.~M Urbas et al.}}},\
  }\bibfield  {title} {\bibinfo {title} {Roadmap on optical metamaterials},\
  }\href {https://doi.org/10.1088/2040-8978/18/9/093005} {\bibfield  {journal}
  {\bibinfo  {journal} {J.\ Opt.}\ }\textbf {\bibinfo {volume} {18}},\ \bibinfo
  {pages} {093005} (\bibinfo {year} {2016})}\BibitemShut {NoStop}%
\bibitem [{\citenamefont {Galfsky}\ \emph {et~al.}(2015)\citenamefont
  {Galfsky}, \citenamefont {Narimanov},\ and\ \citenamefont
  {Menon}}]{Galfsky2015}%
  \BibitemOpen
  \bibfield  {author} {\bibinfo {author} {\bibfnamefont {T.}~\bibnamefont
  {Galfsky}}, \bibinfo {author} {\bibfnamefont {E.~E.}\ \bibnamefont
  {Narimanov}},\ and\ \bibinfo {author} {\bibfnamefont {V.~M.}\ \bibnamefont
  {Menon}},\ }\bibfield  {title} {\bibinfo {title} {Enhanced spontaneous
  emission in photonic hypercrystals},\ }in\ \href
  {https://doi.org/10.1364/FIO.2015.FW6A.3} {\emph {\bibinfo {booktitle}
  {Frontiers in Optics 2015}}},\ \bibinfo {series and number} {OSA Technical
  Digest}\ (\bibinfo {year} {2015})\ \bibinfo {note} {paper FW6A.3, pp. 1,
  2}\BibitemShut {NoStop}%
\bibitem [{\citenamefont {Cortes}\ \emph {et~al.}(2012)\citenamefont {Cortes},
  \citenamefont {Newman}, \citenamefont {Molesky},\ and\ \citenamefont
  {Jacob}}]{Cortesetal2012}%
  \BibitemOpen
  \bibfield  {author} {\bibinfo {author} {\bibfnamefont {C.~L.}\ \bibnamefont
  {Cortes}}, \bibinfo {author} {\bibfnamefont {W.}~\bibnamefont {Newman}},
  \bibinfo {author} {\bibfnamefont {S.}~\bibnamefont {Molesky}},\ and\ \bibinfo
  {author} {\bibfnamefont {Z.}~\bibnamefont {Jacob}},\ }\bibfield  {title}
  {\bibinfo {title} {Quantum nanophotonics using hyperbolic metamaterials},\
  }\href {https://doi.org/10.1088/2040-8978/14/6/063001} {\bibfield  {journal}
  {\bibinfo  {journal} {J.\ Opt.}\ }\textbf {\bibinfo {volume} {14}},\ \bibinfo
  {pages} {063001} (\bibinfo {year} {2012})}\BibitemShut {NoStop}%
\bibitem [{\citenamefont {Cortes}\ \emph {et~al.}(2014)\citenamefont {Cortes},
  \citenamefont {Newman}, \citenamefont {Molesky},\ and\ \citenamefont
  {Jacob}}]{Cortesetal2014-corrigendum}%
  \BibitemOpen
  \bibfield  {author} {\bibinfo {author} {\bibfnamefont {C.~L.}\ \bibnamefont
  {Cortes}}, \bibinfo {author} {\bibfnamefont {W.}~\bibnamefont {Newman}},
  \bibinfo {author} {\bibfnamefont {S.}~\bibnamefont {Molesky}},\ and\ \bibinfo
  {author} {\bibfnamefont {Z.}~\bibnamefont {Jacob}},\ }\bibfield  {title}
  {\bibinfo {title} {Corrigendum: Quantum nanophotonics using hyperbolic
  metamaterials},\ }\href {https://doi.org/10.1088/2040-8978/16/12/129501}
  {\bibfield  {journal} {\bibinfo  {journal} {J.\ Opt.}\ }\textbf {\bibinfo
  {volume} {16}},\ \bibinfo {pages} {129501} (\bibinfo {year}
  {2014})}\BibitemShut {NoStop}%
\bibitem [{\citenamefont {Lucas}\ and\ \citenamefont
  {Fong}(2018)}]{LucasFong2018}%
  \BibitemOpen
  \bibfield  {author} {\bibinfo {author} {\bibfnamefont {A.}~\bibnamefont
  {Lucas}}\ and\ \bibinfo {author} {\bibfnamefont {K.~C.}\ \bibnamefont
  {Fong}},\ }\bibfield  {title} {\bibinfo {title} {Hydrodynamics of electrons
  in graphene},\ }\href {https://doi.org/10.1088/1361-648X/aaa274} {\bibfield
  {journal} {\bibinfo  {journal} {J.\ Phys.: Condens.\ Matter}\ }\textbf
  {\bibinfo {volume} {30}},\ \bibinfo {pages} {053001} (\bibinfo {year}
  {2018})}\BibitemShut {NoStop}%
\bibitem [{\citenamefont {Maier}\ \emph
  {et~al.}(2018{\natexlab{b}})\citenamefont {Maier}, \citenamefont {Margetis},\
  and\ \citenamefont {Luskin}}]{MaierML18}%
  \BibitemOpen
  \bibfield  {author} {\bibinfo {author} {\bibfnamefont {M.}~\bibnamefont
  {Maier}}, \bibinfo {author} {\bibfnamefont {D.}~\bibnamefont {Margetis}},\
  and\ \bibinfo {author} {\bibfnamefont {M.}~\bibnamefont {Luskin}},\
  }\bibfield  {title} {\bibinfo {title} {{Generation of surface
  plasmon-polaritons by edge effects}},\ }\href
  {https://doi.org/https://dx.doi.org/10.4310/CMS.2018.v16.n1.a4} {\bibfield
  {journal} {\bibinfo  {journal} {Commun.\ Math.\ Sci.}\ }\textbf {\bibinfo
  {volume} {16}},\ \bibinfo {pages} {77} (\bibinfo {year}
  {2018}{\natexlab{b}})}\BibitemShut {NoStop}%
\bibitem [{\citenamefont {Davies}\ \emph {et~al.}(2018)\citenamefont {Davies},
  \citenamefont {Patel}, \citenamefont {Xia}, \citenamefont {Herz},\ and\
  \citenamefont {Johnston}}]{Davies2018}%
  \BibitemOpen
  \bibfield  {author} {\bibinfo {author} {\bibfnamefont {C.~L.}\ \bibnamefont
  {Davies}}, \bibinfo {author} {\bibfnamefont {J.~B.}\ \bibnamefont {Patel}},
  \bibinfo {author} {\bibfnamefont {C.~Q.}\ \bibnamefont {Xia}}, \bibinfo
  {author} {\bibfnamefont {L.~M.}\ \bibnamefont {Herz}},\ and\ \bibinfo
  {author} {\bibfnamefont {M.~B.}\ \bibnamefont {Johnston}},\ }\bibfield
  {title} {\bibinfo {title} {Temperature-dependent refractive index of quartz
  at terahertz frequencies},\ }\href
  {https://doi.org/10.1007/s10762-018-0538-7} {\bibfield  {journal} {\bibinfo
  {journal} {J.\ Infrared Millim.\ Te.}\ }\textbf {\bibinfo {volume} {39}},\
  \bibinfo {pages} {1236} (\bibinfo {year} {2018})}\BibitemShut {NoStop}%
\bibitem [{\citenamefont {Chang}\ \emph {et~al.}(2017)\citenamefont {Chang},
  \citenamefont {Zhang}, \citenamefont {Zhang},\ and\ \citenamefont
  {Cui}}]{Chang17}%
  \BibitemOpen
  \bibfield  {author} {\bibinfo {author} {\bibfnamefont {T.}~\bibnamefont
  {Chang}}, \bibinfo {author} {\bibfnamefont {X.}~\bibnamefont {Zhang}},
  \bibinfo {author} {\bibfnamefont {X.}~\bibnamefont {Zhang}},\ and\ \bibinfo
  {author} {\bibfnamefont {H.-L.}\ \bibnamefont {Cui}},\ }\bibfield  {title}
  {\bibinfo {title} {{Accurate determination of dielectric permittivity of
  polymers from 75GHz to 1.6THz using both S-parameters and transmission
  spectroscopy}},\ }\href {https://doi.org/10.1364/AO.56.003287} {\bibfield
  {journal} {\bibinfo  {journal} {Appl.\ Opt.}\ }\textbf {\bibinfo {volume}
  {56}},\ \bibinfo {pages} {3287} (\bibinfo {year} {2017})}\BibitemShut
  {NoStop}%
\bibitem [{\citenamefont {Halevi}\ \emph {et~al.}(1999)\citenamefont {Halevi},
  \citenamefont {Krokhin},\ and\ \citenamefont {Arriaga}}]{Halevi1999}%
  \BibitemOpen
  \bibfield  {author} {\bibinfo {author} {\bibfnamefont {P.}~\bibnamefont
  {Halevi}}, \bibinfo {author} {\bibfnamefont {A.}~\bibnamefont {Krokhin}},\
  and\ \bibinfo {author} {\bibfnamefont {J.}~\bibnamefont {Arriaga}},\
  }\bibfield  {title} {\bibinfo {title} {Photonic crystal optics and
  homogenization of 2d periodic composites},\ }\href@noop {} {\bibfield
  {journal} {\bibinfo  {journal} {Physical review letters}\ }\textbf {\bibinfo
  {volume} {82}},\ \bibinfo {pages} {719} (\bibinfo {year} {1999})}\BibitemShut
  {NoStop}%
\bibitem [{\citenamefont {Krokhin}\ \emph {et~al.}(2010)\citenamefont {Krokhin}
  \emph {et~al.}}]{Krokhin2010}%
  \BibitemOpen
  \bibfield  {author} {\bibinfo {author} {\bibfnamefont {A.~A.}\ \bibnamefont
  {Krokhin}} \emph {et~al.},\ }\bibfield  {title} {\bibinfo {title} {Long-range
  surface plasmons in dielectric-metal-dielectric structure with highly
  anisotropic substrates},\ }\href@noop {} {\bibfield  {journal} {\bibinfo
  {journal} {Physical Review B}\ }\textbf {\bibinfo {volume} {81}},\ \bibinfo
  {pages} {085426} (\bibinfo {year} {2010})}\BibitemShut {NoStop}%
\end{thebibliography}%

\end{document}